%% file: JVLC2018.tex
\documentclass[times]{elsarticle}

\usepackage{booktabs} % For formal tables
\usepackage{forest}
\usepackage{caption}
\usepackage{subcaption}
\usepackage[strings]{underscore}
\usepackage{url}

\graphicspath{{./figures/}}

\journal{Journal of Visual Languages {\&} Computing}

\begin{document}

\begin{frontmatter}

\title{An Experimental Comparison of Map-like Visualisations and Treemaps}

\author[umelb]{Patrick Cheong-Iao Pang}
\ead{mail@patrickpang.net}

\author[umac]{Robert P. Biuk-Aghai\corref{cor}}
\ead{robertb@umac.mo}

\author[umac]{Simon Fong}
\ead{ccfong@umac.mo}

\author[umac]{Yain-Whar Si}
\ead{fstasp@umac.mo}

\cortext[cor]{Corresponding author}

\address[umac]{Department of Computer and Information Science, Faculty of Science and Technology, University of Macau, Avenida da Universidade, Taipa, Macau S.A.R., China}

\address[umelb]{School of Computing and Information Systems, The University of Melbourne, Parkville, VIC 3010, Australia}

\begin{abstract}
Treemaps have been used in information visualisation for over two
decades. They make use of nested filled areas to represent information
hierarchies such as file systems, library catalogues, etc. Recent
years have witnessed the emergence of visualisations that resemble
geographic maps. In this paper we present a study that compares the
performance of one such map-like visualisation with the original two
forms of the treemap, namely nested and non-nested treemaps. Our study employed
a mixed-method evaluation of accuracy, speed and usability (such as the 
ease-of-use and helpfulness of understanding the information). 
We found that accuracy was highest for the map-like visualisations, 
followed by nested treemaps and lastly non-nested treemaps. Task 
performance was fastest for nested treemaps, 
followed by non-nested treemaps, and then map-like visualisations. 
For usability, nested treemaps was considered slightly more 
helpful than map-like visualisations while non-nested performed poorly. 
We conclude that the results regarding
accuracy are promising for the use of map-like visualisations in tasks
involving the visualisation of hierarchical information, while non-nested
treemap are favoured in tasks requiring speed.
\end{abstract}

\begin{keyword}
Cartographic visualisation \sep map-like visualisation \sep treemap \sep experimental evaluation \sep empirical study 
\end{keyword}

\end{frontmatter}

%%%%%%%%%%%%%%%%%%%%%%%%%%%%%%%%%%%%%%%%%%%%%%%%%%%%%%%%%%%%
\section{Introduction}

Information visualisation is a powerful tool for people to understand
information hierarchies such as file systems and library catalogues,
particularly those in which information is buried deep within lower
levels of the hierarchy. It also provides means for a user to leverage
the power of human perception to analyse and reason about the data
\cite{Hall2013}. Not only can visualisation show structure and
relationships inherent in the information, but also summarise and
transform it into a flattened representation that reduces information
overload \cite{Wills2009}. Among the many forms of information
visualisation, treemaps have been widely applied to information
hierarchies. Relatively more recent is the development and application
of map-like visualisations to hierarchical information. These
represent information structures in a form resembling a geographic
map. This paper compares the performance of map-like visualisations
with treemaps in visualising information hierarchies.

The treemap \cite{JS:1991, Shn:1992} and its variants \cite{Bruls2000,
  Wijk1999, Schreck2006, Liang2012} have for many years been popular
in displaying hierarchical and relational data and are found in many
application domains \cite{Tu2007}.

Map-like visualisations \cite{BAY2015, YBA:2015, Pang:2011}, which
have recently received more attention, display hierarchical data and
resemble geographic maps. In this category of visualisation
hierarchical entities are mapped to visual map elements such as
countries, provinces and counties. In this way, the overview and the
organisational relationships of the underlying data is visible in a
map metaphor, which requires no prior training for it to be understood
\cite{BAPS2014, PBY2016}. In addition, in a recent study map-like
visualisations were found to be easy to read and enjoyable to use
\cite{BAY2015}.

Despite the documented advantages of map-like visualisations in
helping users perceive and use the hierarchical data they represent
\cite{BAY2015,PBY2016}, the performance and the usability of such
visualisations remains
unclear. To the best of our knowledge, there is no previous research
comparing map-like visualisations with other forms of
visualisation. Therefore we set out to compare map-like visualisations
with treemaps, one of the most well-established and widely-used forms
of visualisation for hierarchical data. Our comparison focused on
accuracy and speed of task performance. Extending from our previous work
\cite{BPP2017}, we further report on an enhanced statistical analysis,
a test on usability, and a qualitative analysis on the user feedback.
Among the various treemap
algorithms we selected the original two types of treemap proposed by
Johnson and Shneiderman in the first published treemap paper, namely
the nested and non-nested treemap \cite{JS:1991}.

For this research we recruited 40 participants to complete an
evaluation involving the use of three types of visualisation
(non-nested and nested treemap, and map-like) and to identify the
relationship between entities depicted in the visualisation. 
Our results showed that the accuracy was highest for map-like
visualisations, followed by nested treemaps and non-nested treemaps,
and that these findings were statistically significant. However, both
types of treemaps were better at supporting faster task completion
compared with map-like visualisations. Additionally, for usability,
both nested treemaps and map-like visualisations are competent in
helping readers to understand the underlying data. After analysing the
qualitative feedback from our participants, we conclude that the 
results are promising for sing map-like visualisations as an alternative
in representing hierarchical information.

%%%%%%%%%%%%%%%%%%%%%%%%%%%%%%%%%%%%%%%%%%%%%%%%%%%%%%%%%%%%
\section{Related Work}

%%%%%%%%%%%%%%%%%%%%%%%%%%%%%%
\subsection{Treemap Visualisations}

Originally invented by Shneiderman and his colleagues in the 1990s
\cite{JS:1991, Shn:1992}, treemap visualisations have been extensively
investigated and widely used in many application areas. In order to
overcome some of their drawbacks, such as readability, stability and
ordering of blocks, a number of variants of treemaps have been
proposed, including Squarified Treemaps \cite{Bruls2000}, Cushion
Treemaps \cite{Wijk1999} and Ordered and Quantum Treemaps
\cite{Bederson2002}. However, few of these algorithms aim at improving
the application of treemaps to hierarchical data.

\begin{figure}
  \centering
  \includegraphics[width=\columnwidth]{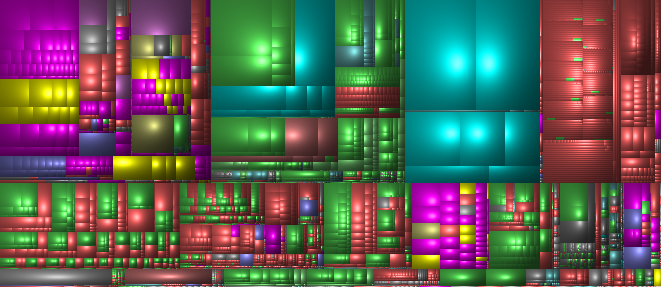}
  \caption{Example of a Cushion Treemap \cite{Kosara} (image licensed
    under CC-BY-SA)}
  \label{fig:image-cushion-treemap}
\end{figure}

\begin{figure}
  \centering
  \includegraphics[width=\columnwidth]{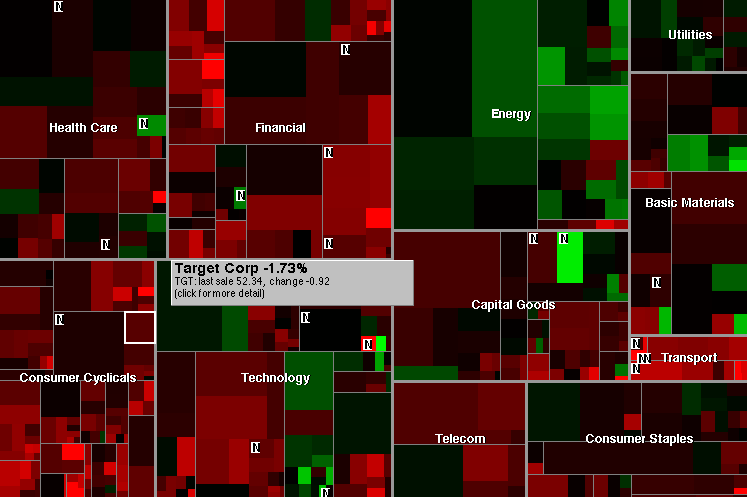}
  \caption{Example of a Squarified Treemap \cite{Kosara} (image
    licensed under CC-BY-SA)}
  \label{fig:image-squarified-treemap}
\end{figure}

An obvious problem of treemaps is that the hierarchical organisation
of data, particularly in cases of deep nesting, cannot be depicted
clearly in the visualisation \cite{Tu2007, Liang2012}. Researchers
have proposed different methods for improving the readability of
nested data in treemaps. For instance, Cushion Treemaps use shadings
and colours to represent areas belonging to the same parent
(Figure~\ref{fig:image-cushion-treemap}); and Squarified Treemaps show
hierarchical data in nested square blocks
(Figure~\ref{fig:image-squarified-treemap}). These approaches allow
the use of treemaps with deeply nested data, and have been adopted to
some actual visualisation problems \cite{Kosara}, including the
visualisation of disk usage that originally motivated the invention of
treemaps. Other approaches for visualising large trees include the use
of non-rectangular areas, such as in Divide and Conquer Treemaps
\cite{LNSH2015}.

Stability and comparability are drawbacks of some treemap algorithms
that are used to visualise hierarchical data. Some small differences
in the hierarchy can cause significant changes to the layout of
treemaps \cite{Hahn2015}, which makes it very hard to compare one
version of the data against another. Some extensions of the treemap
visualisation have been proposed to address this problem. One way is
to limit the aspect ratio of the regions in treemaps
\cite{deBerg2014}, as extreme aspect ratios are difficult to
compare. Another way is to extend the treemap algorithm by using the
properties of Voronoi diagrams, so that changes are only reflected
locally in a smaller area \cite{Hahn2014}. Such work makes treemaps
more useful in applications that require comparison, for example
organisation charts and source code repositories.

%%%%%%%%%%%%%%%%%%%%%%%%%%%%%%
\subsection{Map-like Visualisations}

Map-like visualisations display hierarchical data in the form of a
geographic or topographic map, and are also known as \emph{metaphoric
  maps} \cite{CP:2012}. Cartographic methodologies are employed in
generating these visualisations. A common approach to displaying
hierarchical data with map-like visualisations is to depict data as
cartographic elements (e.g.\ land and sea) and related different types
of data points to map elements \cite{Couclelis1998}. In general,
multiple levels of nested data are conventionally shown as nested
areas in a map (e.g.\ countries, provinces, counties, districts,
etc). As a result lay users can easily perceive the information
contained in the visualisations, as studies have confirmed that most
average readers can effectively read and understand maps in their
daily lives \cite{Borner2003, Borner2010}.

\begin{figure}
  \centering
  \begin{subfigure}[t]{0.3\columnwidth}
    \includegraphics[width=\columnwidth]{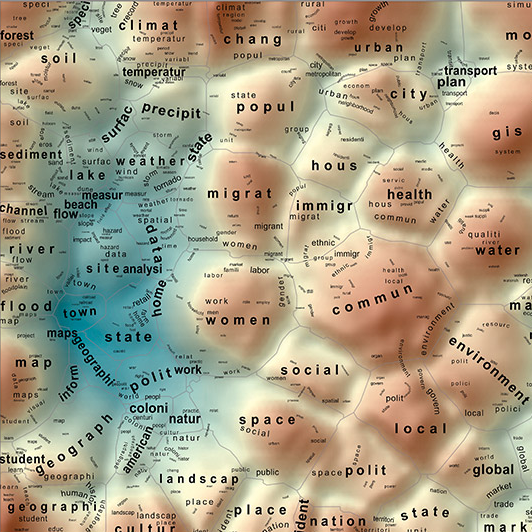}
    \caption{Skupin's SOM map, reproduced from \cite{Skupin2004}}
    \label{fig:skupin}
  \end{subfigure}
  \hfill
  \begin{subfigure}[t]{0.3\columnwidth}
    \includegraphics[width=\columnwidth]{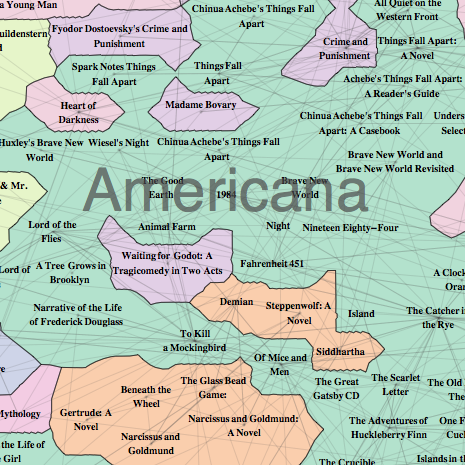}
    \caption{GMap, reproduced from \cite{Mashima2012}}
    \label{fig:gmap}
  \end{subfigure}
  \hfill
  \begin{subfigure}[t]{0.3\columnwidth}
    \includegraphics[width=\columnwidth]{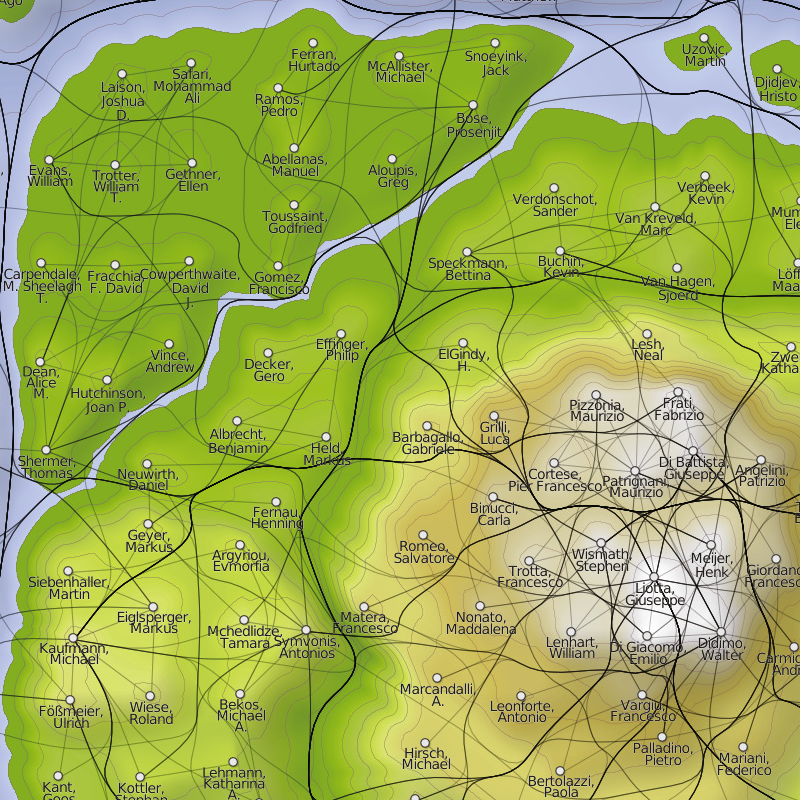}
    \caption{Topological map-like visualisation, reproduced from
      \cite{Gronemann2013}}
    \label{fig:gj}
  \end{subfigure}
  \par\bigskip
  \begin{subfigure}{0.3\columnwidth}
    \includegraphics[width=\columnwidth]{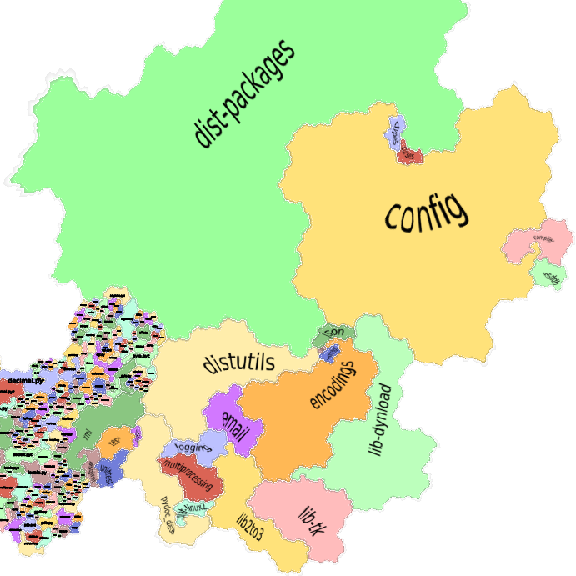}
    \caption{GosperMap, reproduced from \cite{Auber2013}}
    \label{fig:gospermap}
  \end{subfigure}
  \hfill
  \begin{subfigure}{0.3\columnwidth}
    \includegraphics[width=\columnwidth]{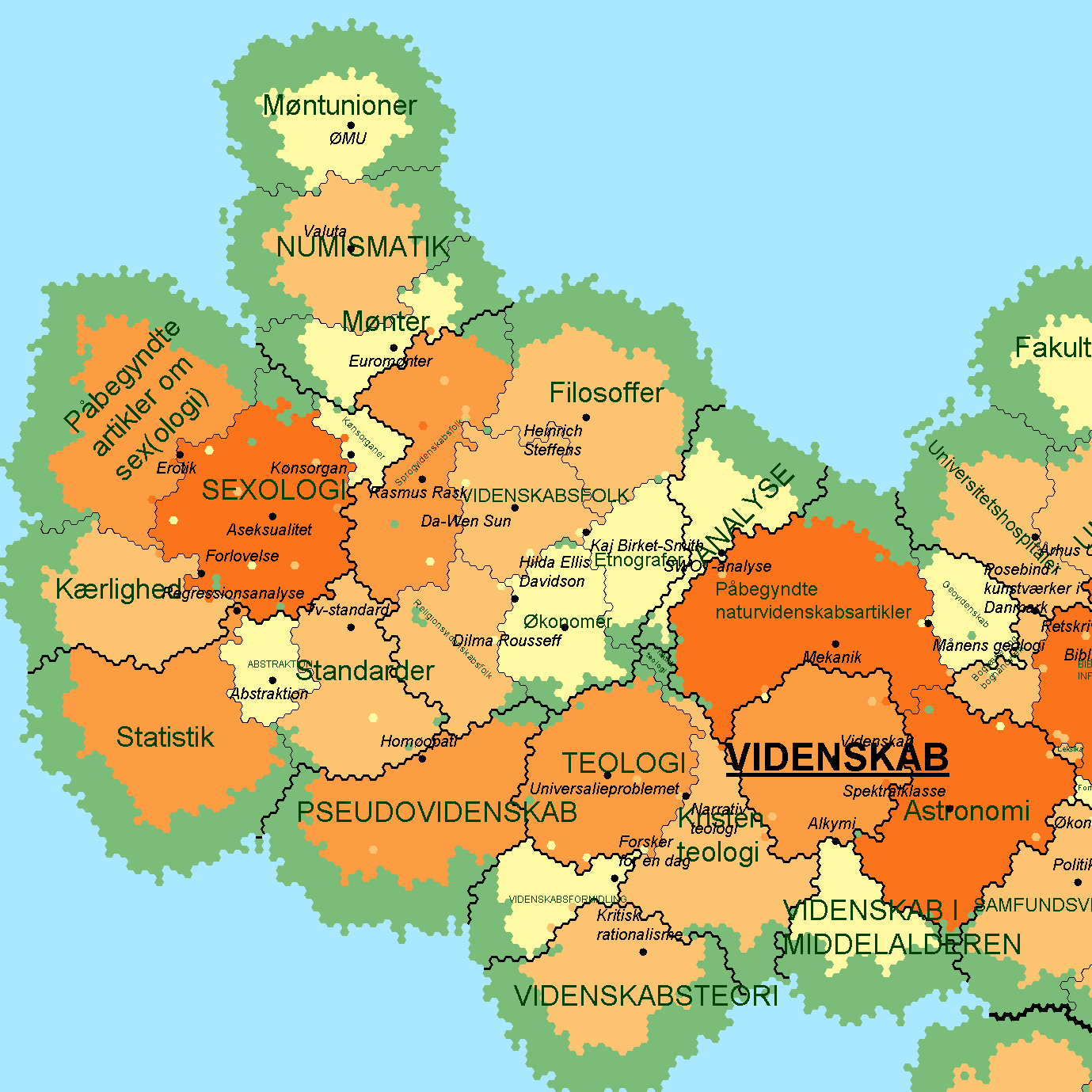}
    \caption{Hexagon tiling-based map-like visualisation, reproduced
      from \cite{BAPS2014}}
    \label{fig:hexagonmap}
  \end{subfigure}
  \hfill
  \begin{subfigure}{0.3\columnwidth}
    \includegraphics[width=\columnwidth]{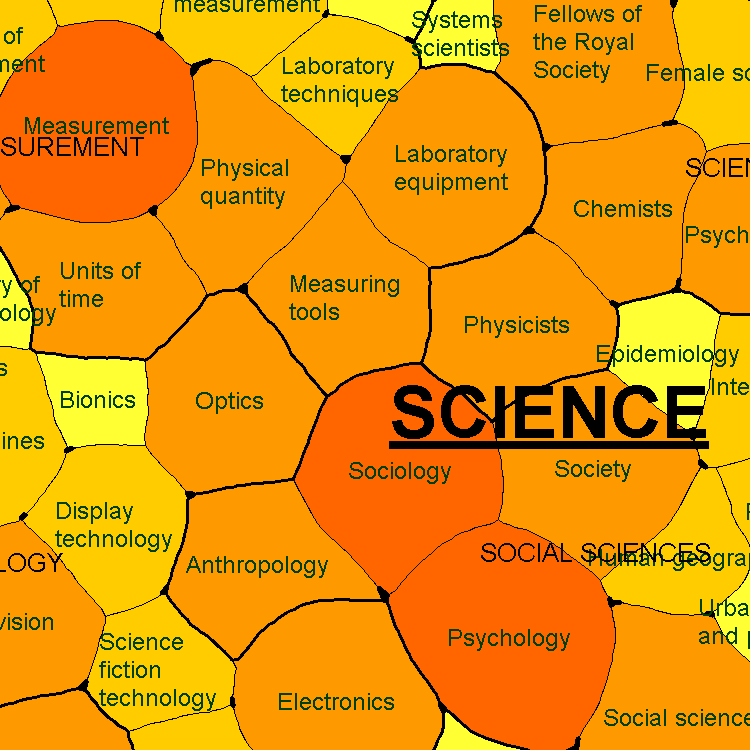}
    \caption{Liquid modelling-based map-like visualisation, reproduced
      from \cite{BAY2015}}
    \label{fig:liquidmap}
  \end{subfigure}
  \label{fig:image-related-work-maplike}
  \par\bigskip
  \begin{subfigure}{1.0\columnwidth}
    \includegraphics[width=\columnwidth]{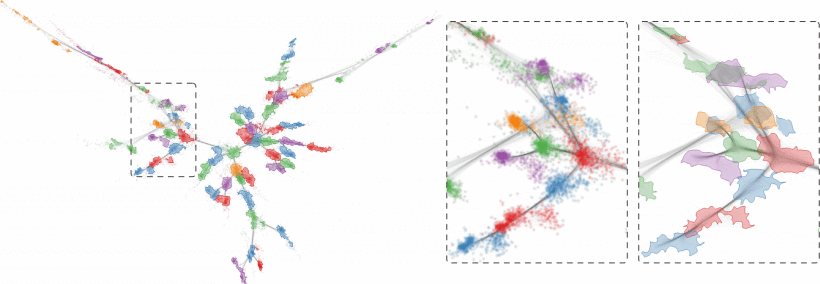}
    \caption{A visualisation with probabilistic graph layout,
      reproduced from \cite{Schulz2017}}
    \label{fig:probgraph}
  \end{subfigure}
  
  \caption{Examples of map-like visualisations}
\end{figure}

A number of algorithms have been proposed for creating map-like
visualisations for different types of data. Skupin suggested using the
self-organising map (SOM) to train a dataset and to visualise the
resulting clusters as land masses (Figure~\ref{fig:skupin})
\cite{Skupin2004}. GMap is another algorithm that draws undirected
graphs as geographic maps (Figure~\ref{fig:gmap}) \cite{Gansner2010,
  Mashima2012}. Gronemann and J{\"u}nger have created an algorithm to
transform networked graphs to topological map-like visualisations
(Figure~\ref{fig:gj}) \cite{Gronemann2013}. Auber et al.\ make use of
Gosper curves for laying out data regions in the visualisation image
(Figure~\ref{fig:gospermap}) \cite{Auber2013}. Recent studies also
highlight the use of hexagons for map-like visualisations by tiling
coloured hexagons on a surface (Figure~\ref{fig:hexagonmap})
\cite{BAPS2014, PBY2016}. Biuk-Aghai et al.\ have applied a liquid
modelling approach to generating map-like visualisations by emulating
the interactions of liquid collisions (Figure~\ref{fig:liquidmap})
\cite{BAY2015}. Although it is not a kind of map-like visualisations,
a recent attempt at visualising data using the probabilistic graph
layout (Figure~\ref{fig:probgraph}) produces output that somewhat
resembles a map \cite{Schulz2017}.

Thanks to the early exposure of maps at an early stage of education
\cite{Blades1998}, they are readily usable by lay users without prior
training. As a result, applications of map-like visualisations are
growing in various domains. For example, they are being used to
illustrate the content and sizes of document corpora \cite{BAPS2014}
and software packages \cite{Auber2013}. Additionally, maps allow
readers to navigate and explore unknown data \cite{Borner2010}, which
is a recent focus in information visualisation. As a result,
visualisations are created to support various information-seeking
tasks, such as browsing, orienteering, exploring and interactive query
refinement \cite{Kairam2015, Khazaei2017}. Some map-like visualisation
applications have the potential for people to discover educational
material \cite{Pang:2011,PVPC2015}, explore knowledge domains \cite{Skupin2004}
and analyse data in the medical context 
\cite{Skupin2013, PHMV2016, PCC2017}. Given this
momentum of extending map-like visualisations to more application
areas, it becomes desirable to understand their effectiveness and how
they compare with existing visualisation alternatives. The
experimental study presented in this paper is one step in this
direction.

%%%%%%%%%%%%%%%%%%%%%%%%%%%%%%%%%%%%%%%%%%%%%%%%%%%%%%%%%%%%
\section{Research Design}

Our research aims to compare a map-like visualisation with the
well-established treemap visualisation, in terms of their ability to
represent hierarchies in an easy-to-understand manner. Specifically,
we are interested to assess our assumption that map-like
visualisations are more intuitive than treemaps when used by lay users
to perform tasks requiring an understanding of hierarchies, and that
this will lead to better performance of a map-like visualisation as
compared to treemaps, both in terms of accuracy and speed. Thus we set
out to test this hypothesis for accuracy (A), presented together with
its null-hypothesis:

\begin{description}
\item [Hypothesis $A_1$:] Map-like visualisations allow lay users to
  perceive hierarchies more accurately than treemaps do.
\item [Hypothesis $A_0$:] Map-like visualisations do not allow lay
  users to perceive hierarchies any more accurately than treemaps do.
\end{description}

Likewise, here is the hypothesis for speed (S), together with its null
hypothesis:

\begin{description}
\item [Hypothesis $S_1$:] Map-like visualisations allow lay users to
  perceive hierarchies faster than treemaps do.
\item [Hypothesis $S_0$:] Map-like visualisations do not allow lay
  users to perceive hierarchies any faster than treemaps do.
\end{description}

To test these hypotheses we designed a task that involved hierarchical
data which we visualised using both a map-like visualisation tool, and
two existing open source treemap visualisation implementations,
producing a nested and a non-nested treemap, respectively. The
hierarchical data set we used was data on student numbers of our
university, arranged in a hierarchy from programme (bachelor, master,
PhD), through academic unit, department, and major, to year. We
collapsed the five levels of this hierarchy into four levels by
aggregating the numbers of students by year to total number of
students over all years.

%%%%%%%%%%%%%%%%%%%%%%%%%%%%%%
\subsection{Visualisation Images}

We created three groups of visualisation images: one group each of
non-nested treemaps, nested treemaps, and map-like visualisations. To
avoid the viewer guessing the right answer from the colours or text
label within the images we labelled areas with meaningless text, and
coloured the background of the entire image in the same colour.

The non-nested treemaps were created using the Protovis software by
the Stanford Visualization
Group \footnote{\url{https://mbostock.github.io/protovis/ex/treemap.html}}. We
modified the software to use only one fill colour for all areas in the
treemap.

The nested treemaps were created using the JavaScript InfoVis Toolkit
by Nicolas Garcia
Belmonte\footnote{\url{https://philogb.github.io/jit/static/v20/Jit/Examples/Treemap/example1.html}}. We
used the squarified tiling algorithm and again modified the software
to only use one fill colour for all area bodies, and one colour for
the area title. This treemap is interactive, so we converted the
produced treemap to a static image for use in our evaluation.

\begin{figure}
  \centering
  \begin{subfigure}{0.53\columnwidth}
    \includegraphics[width=\columnwidth]{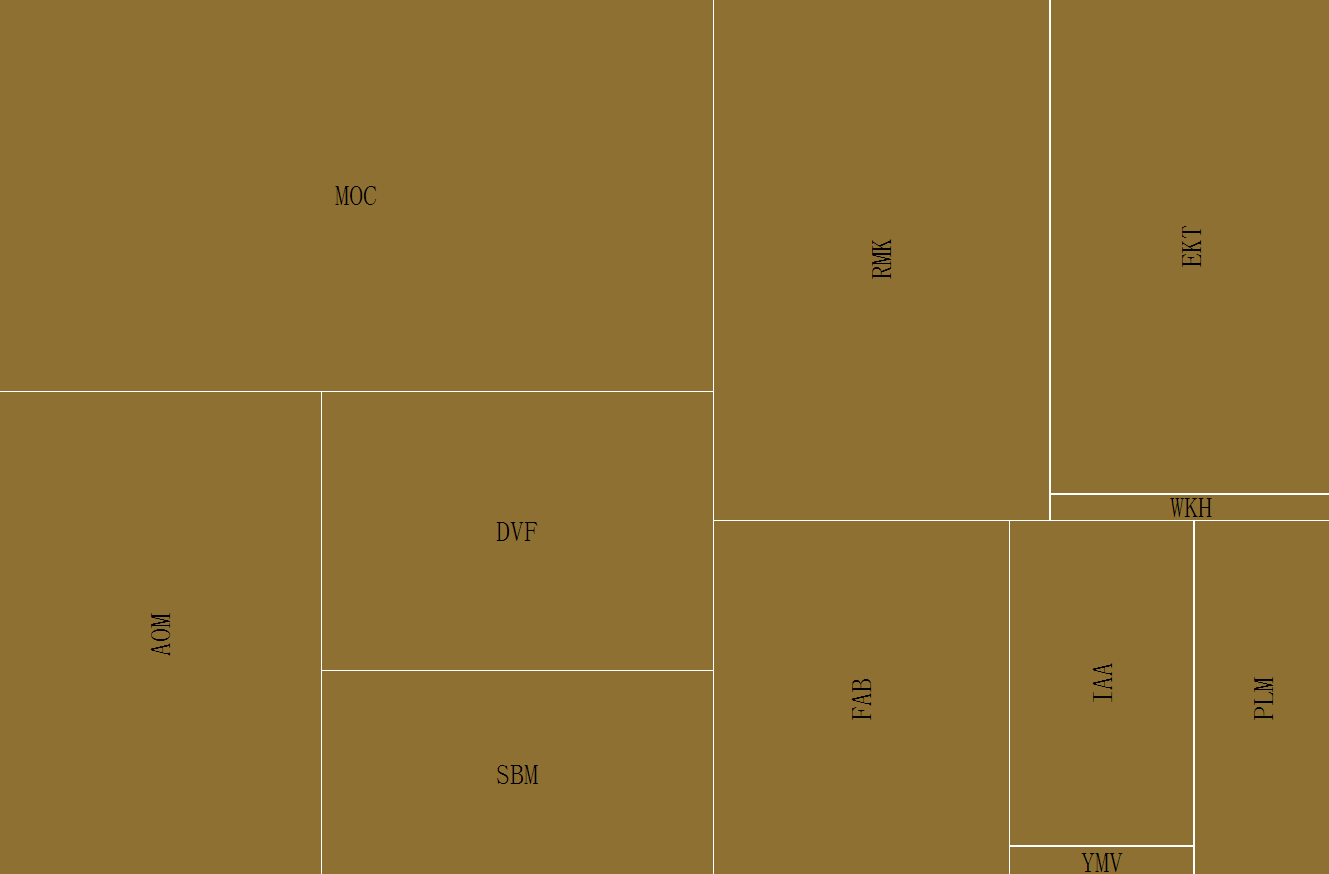}
    \caption{Non-nested treemap}
    \label{fig:non-nested}
  \end{subfigure}
  \par\bigskip
  \begin{subfigure}{0.53\columnwidth}
    \includegraphics[width=\columnwidth]{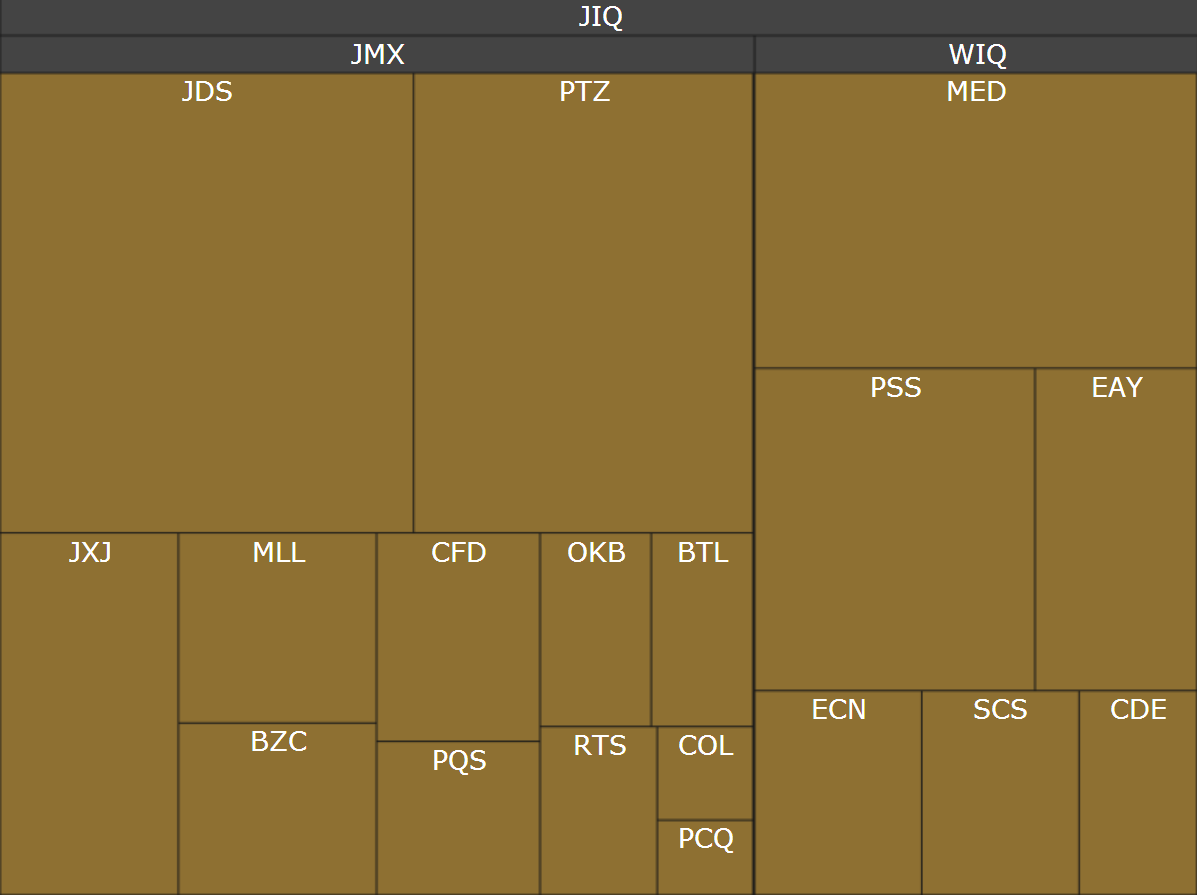}
    \caption{Nested treemap}
    \label{fig:nested}
  \end{subfigure}
  \par\bigskip
  \begin{subfigure}{0.53\columnwidth}
    \includegraphics[width=\columnwidth]{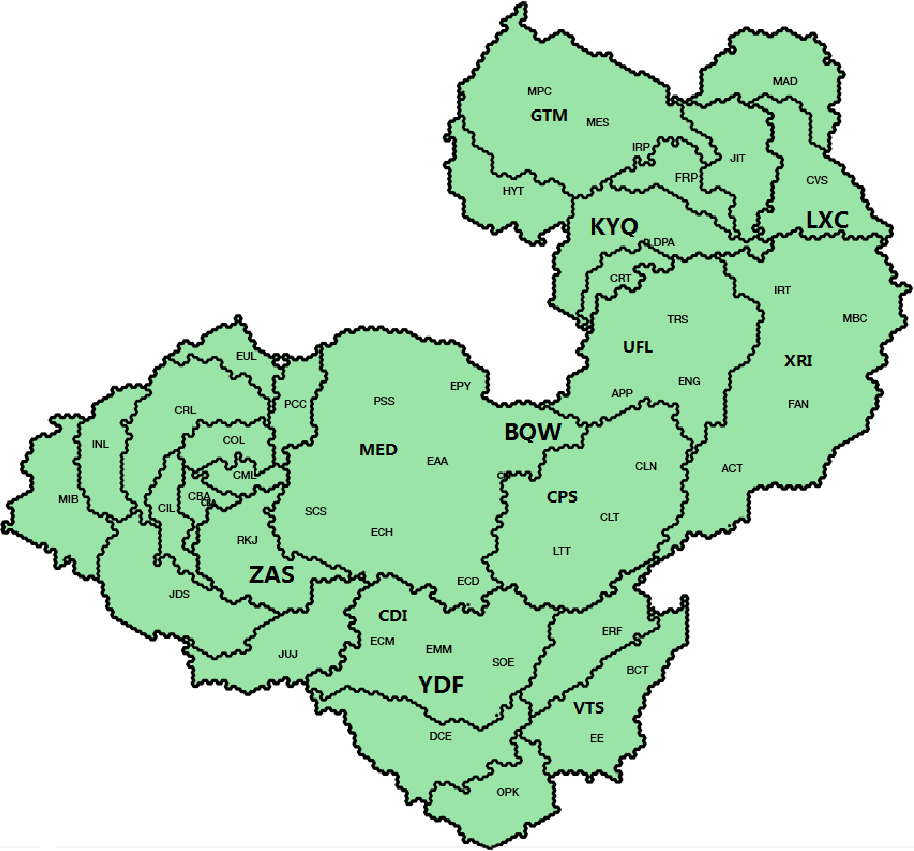}
    \caption{Map-like visualisation}
    \label{fig:map-like}
  \end{subfigure}
  \caption{Samples of visualisation images used in our evaluation}
  \label{fig:image-samples}
\end{figure}

Finally, the map-like visualisation images were created using a
software developed by us that implements the enhanced hexagon tiling
algorithm (EHTA) \cite{YBA:2015}. We chose this particular algorithm
for two reasons: firstly, we have access to the program code, enabling
us to generate map-like visualisations of our data; and secondly, in a
previous study this particular algorithm was found to produce
visualisations that most strongly resemble geographic
maps~\cite{PBY2016}. Samples of the images used are shown in
Figure~\ref{fig:image-samples}.

%%%%%%%%%%%%%%%%%%%%%%%%%%%%%%
\subsection{Survey Design}

The survey was conducted in a controlled setting, in a computer lab in
our university. The researchers gave a brief introduction about the
survey, explaining the concepts of hierarchy and of information
visualisation, and introducing the treemap visualisation for
representing hierarchies. This was followed by the survey itself which
was conducted by each participant at a computer, accessing a survey
website prepared by us containing the visualisation images described
above. This online survey consisted of five parts:

\begin{enumerate}
\item Informed consent form: explaining that participation is
  voluntary and requiring participant agreement to continue with the
  study (as required by our university's ethics committee).

\item Entry questionnaire: collecting information on the participant's
  age, gender, degree pursued (bachelor, master or PhD), IT skills (on
  a 7-point Likert scale ranging from ``know nothing'' to ``know how
  to write computer programs''), and knowledge about information
  visualisation (on a 7-point Likert scale ranging from ``not at all''
  to ``expert'').

\item Practice: asking the participant to evaluate three visualisation
  images and to answer questions about the relationship of areas in
  the image. The evaluation task is explained in more detail
  below. The purpose of the practice questions was to serve as a
  warm-up to let participants get used to the questions to be answered
  in the following main evaluation part of the survey. However, there
  was no indication in the user interface that this part was a
  practice only, so participants performed this task as if it counted
  toward the actual evaluation. Answers collected in this part were
  not evaluated.

\item Evaluation: this is the main part of the survey in which the
  participant answered questions about the visualisation images, in
  different order by group as defined below. On each page, five images
  were presented to the participant and for each image the participant
  was asked to answer one multiple-choice question as explained below.

\item Exit questionnaire: here we revealed to the participant what the
  visualisation images represented, namely our university's student
  enrolment numbers, and asked for an overall assessment of the
  visualisation. The overall assessment included the perceived levels of
  ease-of-use and helpfulness of the visualisations for the given tasks, 
  as well as an open-ended question for collecting qualitative
  feedback about the visualisations. For qualitative comments, we adopt
  an open coding approach \cite{saldana2015coding} to obtain important
  insights from the data.
\end{enumerate}

To eliminate the effect of ordering on results we divided participants
in four groups, each of which evaluated the same sets of images but in
different order. The basic ordering was treemap vs.\ map-like
visualisation images, and within the group of treemap images a further
ordering was made with nested vs. non-nested images. The order of
images evaluated by the four groups of participants were thus as
follows:

\begin{description}
\item [Group A:] non-nested $\rightarrow$ nested $\rightarrow$
  map-like
\item [Group B:] nested $\rightarrow$ non-nested $\rightarrow$
  map-like
\item [Group C:] map-like $\rightarrow$ non-nested $\rightarrow$
  nested
\item [Group D:] map-like $\rightarrow$ nested $\rightarrow$
  non-nested
\end{description}

%%%%%%%%%%%%%%%%%%%%%%%%%%%%%%
\subsection{Evaluation Task}

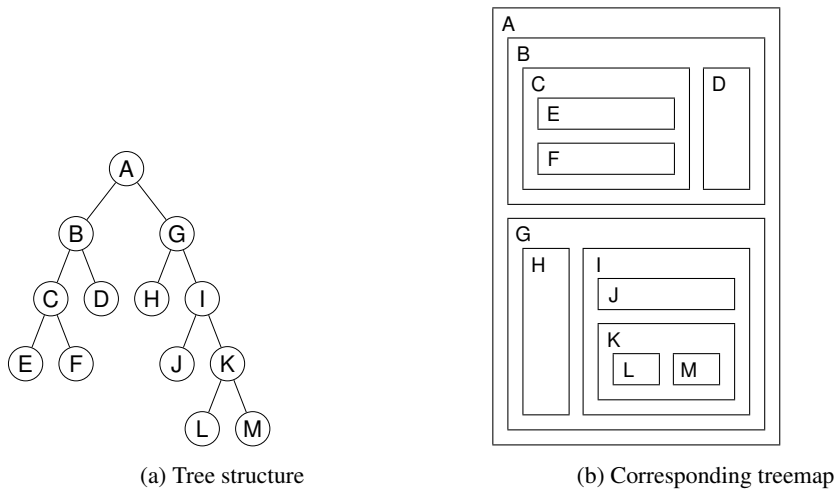
\begin{figure}
  \centering
  \begin{subfigure}[b]{0.47\columnwidth}
    \scalebox{0.91}{\sffamily
      \begin{forest}
        for tree={circle, draw, 
          minimum size=1.5em,
          inner sep=1pt}
        [A
          [B
            [C
              [E]
              [F]]
            [D]]
          [G
            [H]
            [I
              [J]
              [K
                [L]
                [M]]]]]
      \end{forest}
    }
    \caption{Tree structure}
    \label{fig:example-tree}
  \end{subfigure}
  \hfill
  \begin{subfigure}[b]{0.47\columnwidth}
    \input{figures/treemap-sample.tex}
    \caption{Corresponding treemap}
    \label{fig:example-treemap}
  \end{subfigure}
  \caption{Example hierarchy}
  \label{fig:example-hierarchy}
\end{figure}

The evaluation task required participants to recognise the
relationship between two areas shown in the visualisation. As areas in
the visualisation represent nodes in a hierarchy, this means that the
task involved recognising this relationship between the
nodes. Figure~\ref{fig:example-hierarchy} shows an example
illustrating this. Figure~\ref{fig:example-tree} depicts the structure
of a hierarchy, which is a rooted tree. Each node other than the root
has a parent node, which is the node immediately above it in the tree,
linked to it by an edge; and it may have multiple child nodes, which
are nodes below it in the tree, linked to it by edges. We name the
parent directly connected to a node its \emph{direct parent}, and
likewise there may be multiple \emph{direct child} nodes. For example,
the direct parent of node {\sffamily I} is node {\sffamily G}, and the
direct children of node {\sffamily I} are nodes {\sffamily J} and
{\sffamily K}. The parent of a parent is called an \emph{indirect
  parent}, and likewise the indirect parent of a parent recursively
all the way up to the root. The child of a child is called an
\emph{indirect child}, and likewise the indirect child of a child
recursively down the tree. For example, node {\sffamily M} has the
direct parent {\sffamily K} and the indirect parents {\sffamily I},
{\sffamily G} and {\sffamily A}. Node {\sffamily B} has the direct
children {\sffamily C} and {\sffamily D}, and the indirect children
{\sffamily E} and {\sffamily F}.

The tree structure of Figure~\ref{fig:example-tree} is represented as
the treemap shown in Figure~\ref{fig:example-treemap}. In the treemap,
parent-child relationships (direct and indirect) are represented
through the nesting of areas: parents contain their direct and
indirect children, recursively down the hierarchy; conversely,
children are contained within their direct and indirect parents,
recursively up the hierarchy. For example, in
Figure~\ref{fig:example-treemap} area {\sffamily M} is contained
within area {\sffamily K} (its direct parent), which is contained
within {\sffamily I}, which is contained within {\sffamily G}, which
is contained within {\sffamily A}, these being its indirect
parents. Similarly, child relationships are represented in the same
way.

A task involving the recognition of the hierarchical relationship
within a treemap could for example ask what the relationship between
areas {\sffamily M} and {\sffamily K} is (correct answer: {\sffamily
  M} is a direct child of {\sffamily K}); or what the relationship
between areas {\sffamily G} and {\sffamily K} is (correct answer:
{\sffamily G} is an indirect parent of {\sffamily K}); or what the
relationship between areas {\sffamily C} and {\sffamily D} is (correct
answer: {\sffamily C} and {\sffamily D} are at the same level).

Our evaluation task presented an image such as one of those shown in
Figure~\ref{fig:image-samples} and asked the participant following
question:

\begin{quote}
  What is the relationship between {\sffamily A1} and {\sffamily A2}?
\end{quote}

The possible answers provided were:

\begin{quote}
  \begin{enumerate}
  \item {\sffamily A1} is an indirect parent of {\sffamily A2}

  \item {\sffamily A1} is a direct parent of {\sffamily A2}

  \item {\sffamily A1} is an indirect child of {\sffamily A2}

  \item {\sffamily A1} is a direct child of {\sffamily A2}

  \item {\sffamily A1} and {\sffamily A2} are at the same level

  \item I don't know
  \end{enumerate}
\end{quote}

In the actual evaluation, instead of {\sffamily A1} and {\sffamily
  A2}, the question and answer text showed the names of areas existing
in the image, such as {\sffamily MED} and {\sffamily PSS} in
Figure~\ref{fig:map-like} (in this case the former is a direct parent
of the latter). The terms direct parent, indirect parent, direct child
and indirect child had been introduced during the brief introduction
prior to the beginning of the survey, so participants were familiar
with the meaning of these terms in the context of our evaluation.

%%%%%%%%%%%%%%%%%%%%%%%%%%%%%%
\subsection{Participant Recruitment}

We invited students to join our survey through our university's
student associations, who sent out invitation messages through social
media (Facebook and WeChat). These messages reached thousands of
students from across all our university's academic units, majors, and
degree programmes. We believe this recruitment process helped ensure
that a representative sample of students was recruited. Our invitation
message asked students to participate in our research in one of the
computer labs on campus, that it would take about 1 hour, and that
each student would be rewarded for their participation with a
supermarket coupon (of about USD6.25 value). 60 students signed up for
our survey, and finally 40 students participated.

%%%%%%%%%%%%%%%%%%%%%%%%%%%%%%%%%%%%%%%%%%%%%%%%%%%%%%%%%%%%
\section{Results and Discussion}

We present results of the demographic survey, of accuracy and speed of
task performance. Also, we report on the results of the exit questionnaire
regarding to the perceived ease-of-use and helpfulness for understanding 
the data, as well as the qualitative feedback. In addition, we include a
discussion of the results.

%%%%%%%%%%%%%%%%%%%%%%%%%%%%%%
\subsection{Demographic Results}

\begin{table}
  \centering
  \caption{Degree Pursued by Participants}
  \label{tab:degree}
  \small
  \begin{tabular}{lrr} \toprule
    \textbf{Degree}   & \textbf{N} & \textbf{\%} \\ \midrule
    Bachelor   & 14  & 35.0\% \\
    Master     & 21  & 52.5\% \\
    PhD        &  5  & 12.5\% \\ \bottomrule
  \end{tabular}
\end{table}

\begin{table}
  \centering
  \caption{Academic Unit of Participants}
  \label{tab:acadunit}
  \small
  \begin{tabular}{lrr} \toprule
    \textbf{Academic Unit}   & \textbf{N} & \textbf{\%} \\ \midrule
    Arts \& Humanities      & 6  & 15.0\% \\
    Business Administration & 6  & 15.0\% \\
    Chinese Medical Science & 2  &  5.0\% \\
    Education               & 1  &  2.5\% \\
    Law                     & 4  & 10.0\% \\
    Social Science          & 7  & 17.5\% \\
    Science \& Technology   & 14 & 35.0\% \\ \bottomrule
  \end{tabular}
\end{table}

A total of 40 participants completed the survey, equally divided into
Groups A, B, C, and D as explained above, i.e.\ with 10 participants
per group. There were 22 female and 18 male students, with a mean age
of 22.6 years and a median age of 23 years. 14 of them were Bachelor
students, 21 were Master students and 5 were PhD students
(representing 35.0\%, 52.5\% and 12.5\% of participants,
respectively).  The degree pursued by these students is summarised in
Table~\ref{tab:degree} and our participants came from almost all academic
disciplines, with 14 (35\%) from science and technology, and 26 (65\%)
from disciplines such as business, education, law, social sciences and
others.  their distribution across academic units is shown in
Table~\ref{tab:acadunit}.  Participants self-assessed their IT skills
ranging from a lowest value of 3 (``know how to use office software
and Internet'') to a highest value of 7 (``know how to write computer
programs''), with a mean value of 5.2, higher than the mid-point value
of 4. In terms of knowledge of visualisation they assessed themselves
ranging from a low of 1 (``not at all'') to 6 (between ``neutral'' and
``expert''), with a mean value of 3.6 which is somewhat lower than the
mid-point value of 4. Thus this sample of participants was
technologically adept, but in terms of knowledge about visualisation
could be considered lay users, which is what we assume for our
hypotheses.

%%%%%%%%%%%%%%%%%%%%%%%%%%%%%%
\subsection{Accuracy}

\begin{figure}
  \centering
  \includegraphics[width=\columnwidth]{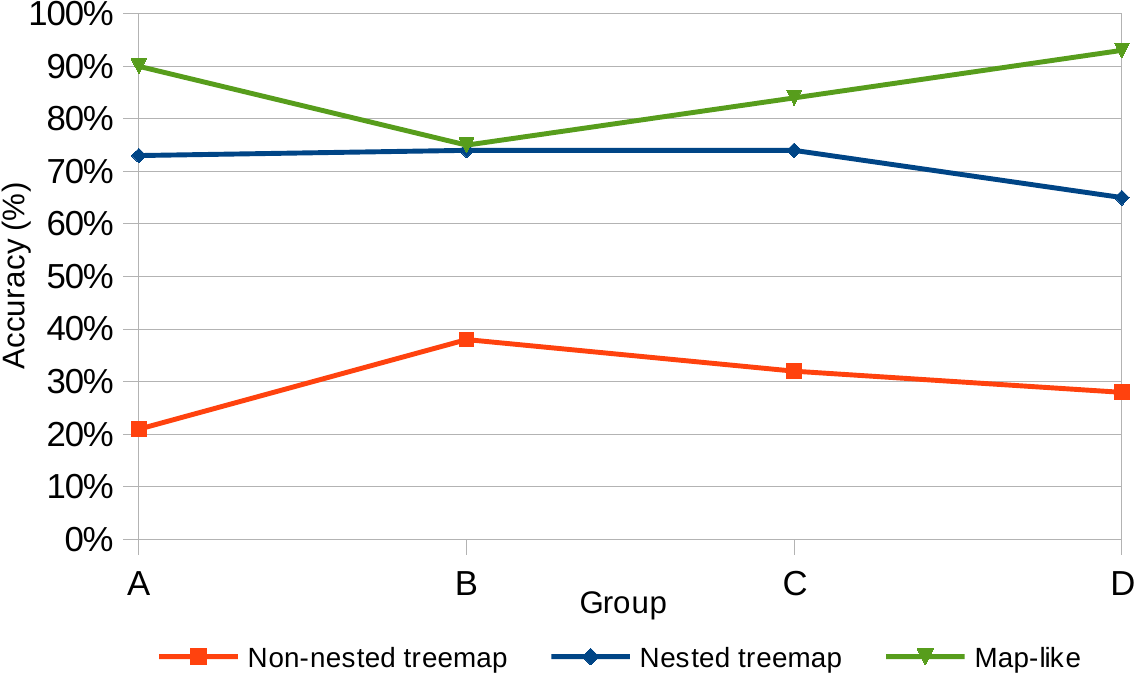}
  \caption{Mean accuracy by group}
  \label{fig:acc-group}
\end{figure}

We collected responses to the evaluation questions and compared these
with the correct answer. Each answer was then mapped to a binary
true--false value, i.e.\ answers were considered either completely
right or completely wrong.
Tables~\ref{tab:acc-nntm}, \ref{tab:acc-ntm} and \ref{tab:acc-mlv}
show the summary of results for accuracy for non-nested treemaps,
nested treemaps, and map-like visualisations, respectively.
Table~\ref{tab:acc-group} shows the mean values of accuracy for each
of the four groups and all three types of visualisation, and
Figure~\ref{fig:acc-group} shows a plot of these accuracy values.
i.e.\ the values for groups A \ldots\ D in the last column of
Tables~\ref{tab:acc-nntm}, \ref{tab:acc-ntm} and \ref{tab:acc-mlv}.

\begin{table*}
  \centering
  \caption{Mean accuracy by group for non-nested treemaps (images NN1
    \ldots\ NN10)}
  \label{tab:acc-nntm}
  \footnotesize
  \begin{tabular}{crrrrrrrrrrr} \toprule
    \textbf{Group} & \textbf{NN1} & \textbf{NN2} & \textbf{NN3} & \textbf{NN4} & \textbf{NN5} & \textbf{NN6} & \textbf{NN7} & \textbf{NN8} & \textbf{NN9} & \textbf{NN10} & \textbf{Avg.} \\ \midrule
    A & 10\% & 10\% & 20\% & 10\% & 20\% & 20\% & 40\% & 40\% & 0\% & 40\% & 21\% \\
    B & 10\% & 10\% & 60\% & 50\% & 20\% & 50\% & 70\% & 40\% & 30\% & 40\% & 38\% \\
    C & 20\% & 20\% & 40\% & 50\% & 30\% & 10\% & 90\% & 30\% & 10\% & 20\% & 32\% \\
    D & 10\% & 10\% & 10\% & 40\% & 10\% & 20\% & 60\% & 60\% & 10\% & 50\% & 28\% \\ \midrule
    Avg. & 13\% & 13\% & 33\% & 38\% & 20\% & 25\% & 65\% & 43\% & 13\% & 38\% & 30\% \\ \bottomrule
  \end{tabular}
\end{table*}

\begin{table*}
  \centering
  \caption{Mean accuracy by group for nested treemaps (images N1 \ldots\ N10)} %
  \label{tab:acc-ntm}
  \footnotesize
  \begin{tabular}{crrrrrrrrrrr} \toprule
    \textbf{Group} & \textbf{N1} & \textbf{N2} & \textbf{N3} & \textbf{N4} & \textbf{N5} & \textbf{N6} & \textbf{N7} & \textbf{N8} & \textbf{N9} & \textbf{N10} & \textbf{Avg.} \\ \midrule
    A & 70\% & 70\% & 90\% & 40\% & 40\% & 70\% & 90\% & 70\% & 90\% & 100\% & 73\% \\ 
    B & 70\% & 30\% & 80\% & 40\% & 60\% & 80\% & 90\% & 100\% & 90\% & 100\% & 74\% \\ 
    C & 80\% & 40\% & 80\% & 30\% & 70\% & 100\% & 90\% & 80\% & 70\% & 100\% & 74\% \\ 
    D & 50\% & 70\% & 40\% & 70\% & 70\% & 50\% & 90\% & 60\% & 70\% & 80\% & 65\% \\ \midrule
    Avg. & 68\% & 53\% & 73\% & 45\% & 60\% & 75\% & 90\% & 78\% & 80\% & 95\% & 72\% \\ \bottomrule
  \end{tabular}
\end{table*}

\begin{table*}
  \centering
  \caption{Mean accuracy by group for map-like visualisation (images
    M1 \ldots\ M10)}
  \label{tab:acc-mlv}
  \scriptsize
  \begin{tabular}{crrrrrrrrrrr} \toprule
    \textbf{Group} & \textbf{M1} & \textbf{M2} & \textbf{M3} & \textbf{M4} & \textbf{M5} & \textbf{M6} & \textbf{M7} & \textbf{M8} & \textbf{M9} & \textbf{M10} & \textbf{Avg.} \\ \midrule
    A & 40\% & 80\% & 100\% & 100\% & 100\% & 80\% & 100\% & 100\% & 100\% & 100\% & 90\% \\ 
    B & 20\% & 90\% & 90\% & 80\% & 100\% & 30\% & 100\% & 80\% & 80\% & 80\% & 75\% \\ 
    C & 80\% & 90\% & 70\% & 80\% & 90\% & 70\% & 90\% & 80\% & 90\% & 100\% & 84\% \\ 
    D & 80\% & 90\% & 90\% & 100\% & 100\% & 90\% & 100\% & 80\% & 100\% & 100\% & 93\% \\ \midrule
    Avg. & 55\% & 88\% & 88\% & 90\% & 98\% & 68\% & 98\% & 85\% & 93\% & 95\% & 86\% \\ \bottomrule
  \end{tabular}
\end{table*}

\begin{table}
  \centering
  \caption{Mean accuracy by group for non-nested treemaps, nested
    treemaps and map-like visualisations (ranges of the 95\%
    confidence interval in parentheses)}
  \label{tab:acc-group}
  \small
  \begin{tabular}{crrrr} \toprule
    \textbf{Group} & \textbf{Non-nested} & \textbf{Nested} & \textbf{Map-like} & \textbf{Avg.} \\ \midrule
    A & 21\% ($\pm$ 12.9\%) & 73\% ($\pm$ 10.6\%) & 90\% ($\pm$5.1\%) & 61\% \\ 
    B & 38\% ($\pm$ 11.6\%) & 74\% ($\pm$ 13.1\%) & 75\% ($\pm$ 13.8\%) & 62\% \\ 
    C & 32\% ($\pm$ 10.9\%) & 74\% ($\pm$ 11.8\%) & 84\% ($\pm$ 16.1\%) & 63\% \\ 
    D & 28\% ($\pm$ 10.0\%) & 65\% ($\pm$ 18.3\%) & 93\% ($\pm$ 7.2\%) & 62\% \\ \midrule
    Avg. & 30\% & 72\% & 86\% & 62\% \\ \bottomrule
  \end{tabular}
\end{table}

The results for accuracy show that for all groups the accuracy of the
map-like visualisation was highest, ranging between 75\% and 93\% mean
accuracy per group, followed by the nested treemap (ranging between
65\% and 74\%), and the non-nested treemap finishing a distant third
(21\% to 38\%). The gap in accuracy between the highest and second
highest visualisation per group was noticeable for most groups, by up
to 28 percentage points (group D). Only group B had almost the same
accuracy score for the map-like and nested treemap visualisations. The
gap between the second and third highest scoring visualisation was
even greater, ranging between 36 and 52 percentage
points. Interestingly, despite these differences in accuracy, the
average accuracy per group across the three types of visualisation
evaluated was almost identical, ranging from 61\% to 63\%.

Within each group and for each image we counted the number of
visualisation images with a 100\% accuracy score, i.e.\ where each of
the 10 participants in a group got the answer right. There were three
times as many such perfect scores for the map-like visualisation than
for the nested treemap (15 vs.\ 5 out of 40 scores per type of
visualisation). The non-nested treemap did not have any scores of
100\%, reaching as its highest a score of 90\% which occurred only
once. On the contrary, the non-nested treemap was the only
visualisation that had a score of 0\% for one of the images and
groups, something that none of the other visualisations encountered.

We wondered whether inter-group differences in accuracy could be
attributed to a learning effect, however the data did not support this
assumption. Groups A and B first evaluated treemaps and then the
map-like visualisation, whereas groups C and D evaluated the map-like
visualisation first and then treemaps. If a learning effect had been
present then the later evaluations would have benefited from the
experience of having performed the earlier tasks. The same task
performed later would thus have achieved a higher accuracy score than
for those groups in which the order was the reverse. Observing the
mean accuracy values between groups, however, we observe that this is
not actually the case: groups C and D used the map-like visualisation
first, whereas groups A and B used the map-like visualisation
later. If groups A and B would have learned to perform the task better
from their experience of having done similar tasks on treemaps, then
the accuracy scores for map-like visualisations in groups A and B
should be higher than in groups C and D. But the mean accuracy score
of groups A and B combined is 82.5\%, whereas the corresponding score
for groups C and D combined is 88.5\%. Comparing nested treemaps
evaluated first or later we can see that only group B has evaluated
nested treemaps first, but has the highest accuracy score (74\%, the
same as group C where it was evaluated last). In groups A and D nested
treemaps were evaluated second, but score quite differently. Thus
there does not appear to be a learning effect due to the order of
evaluation of different visualisations.

\begin{figure}
  \centering
  \includegraphics[width=\columnwidth]{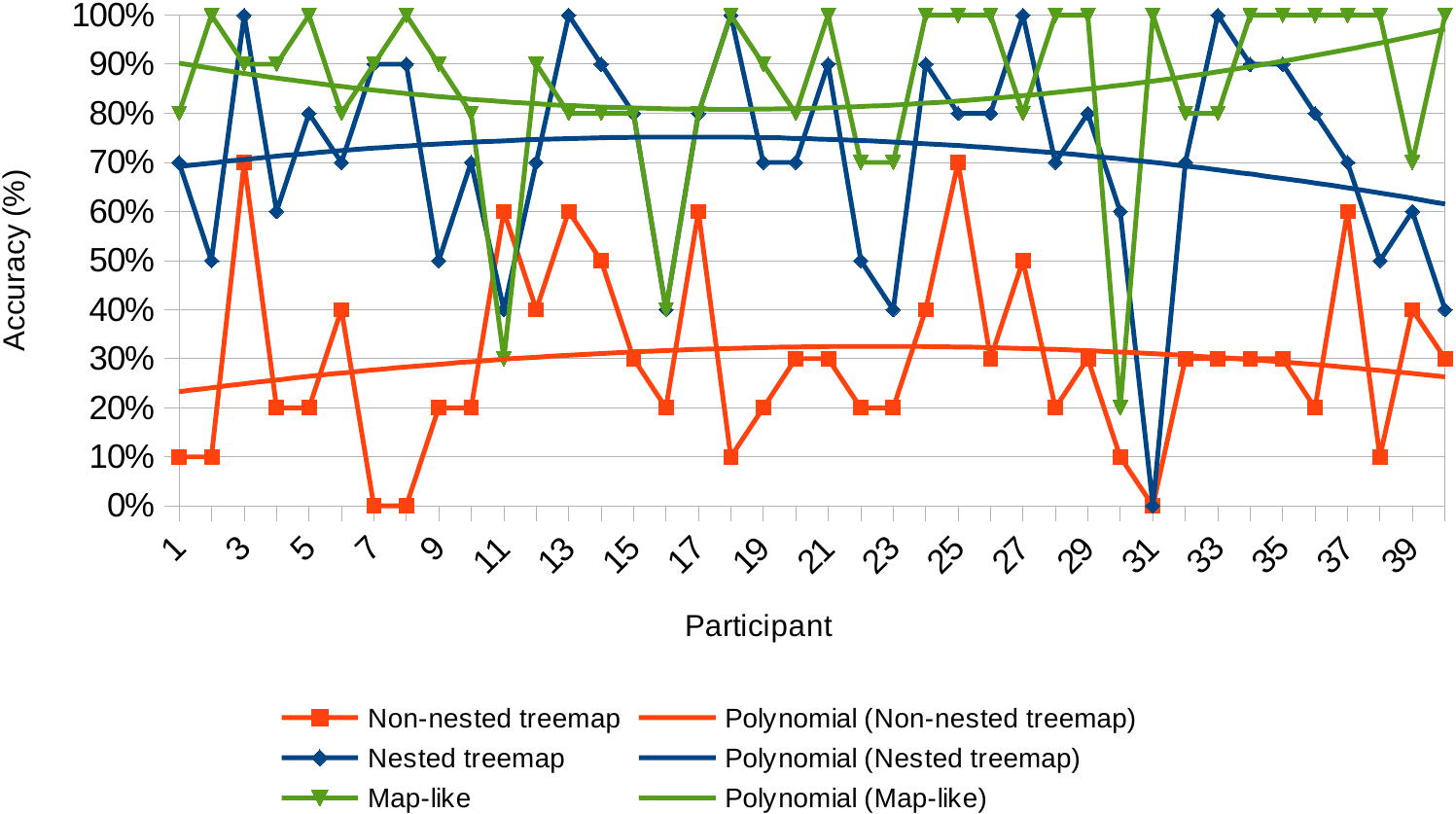}
  \caption{Accuracy by individual participants}
  \label{fig:acc-participant}
\end{figure}

To determine whether the difference in accuracy between the different
kinds of visualisations was statistically significant we decided to
look at performance on the level of individual participants, not at
the group level. This is because we observed that there existed large
variations in individual performance within a group. Thus we compared
the performance in terms of accuracy of individual participants across
the different visualisations. Figure~\ref{fig:acc-participant} shows
accuracy of each of the 40 participants (the first ten participants
belonged to group A, the next ten to group B, and so on). We can
notice a great amount of fluctuation: for non-nested treemaps it
ranges through 70\% (0\%--70\%), for nested treemaps the range is
100\% (0\%--100\%), and for the map-like visualisation the range is
80\% (20\%--100\%). We overlayed a polynomial trendline to smoothen
these fluctuations and perceive a pattern. These trendlines again show
the great distance between the non-nested and nested treemap accuracy
figures, and a smaller but still clear distance between the nested
treemap and map-like visualisation figures. The distance was smallest
around participant 17, who belongs to group B, echoing our observation
from the result of groups above. Detailed accuracy values of
individual participants for each type of visualisation are shown in
Table~\ref{tab:acc-participant} in \ref{app:individual-responses}.

We performed a two-tailed paired t-test on the accuracy figures,
comparing pairs of visualisations that were neighbours in terms of
accuracy. Additionally, an ANOVA test was performed to reassure the 
significant differences seen among different visualisations.
The results are presented in Table~\ref{tab:acc-significance}.
The t-tests indicate that the difference in
accuracy between non-nested and nested treemaps was highly significant
($p$ close to zero), as was the difference between nested treemaps and
map-like visualisations ($p < 0.001$). The ANOVA test shows a significant 
differences among the means of accuracy across three visualisations 
($p < 0.001, F(2, 117) = 85.216, \etaup^2 = 0.593$). 
Given these results we can thus
conclude that map-like visualisations indeed allow more accurate
perception of hierarchy, which allows us to accept Hypothesis $A_1$.

\begin{table}
  \centering
  \caption{Statistical significance of accuracy differences by
    individual participants}
  \label{tab:acc-significance}
  \small
  \begin{tabular}{lr} \toprule
    \textbf{Statistical Test} & \textbf{$p$-value} \\ \midrule
    Non-nested vs.\ nested treemap & 8.765E-14 \\
    Nested treemap vs.\ map-like   & 0.0008 \\ 
    ANOVA & 1.461E-23 \\ \bottomrule
  \end{tabular}
\end{table}

%%%%%%%%%%%%%%%%%%%%%%%%%%%%%%
\subsection{Speed}

\begin{figure}
  \centering
  \includegraphics[width=\columnwidth]{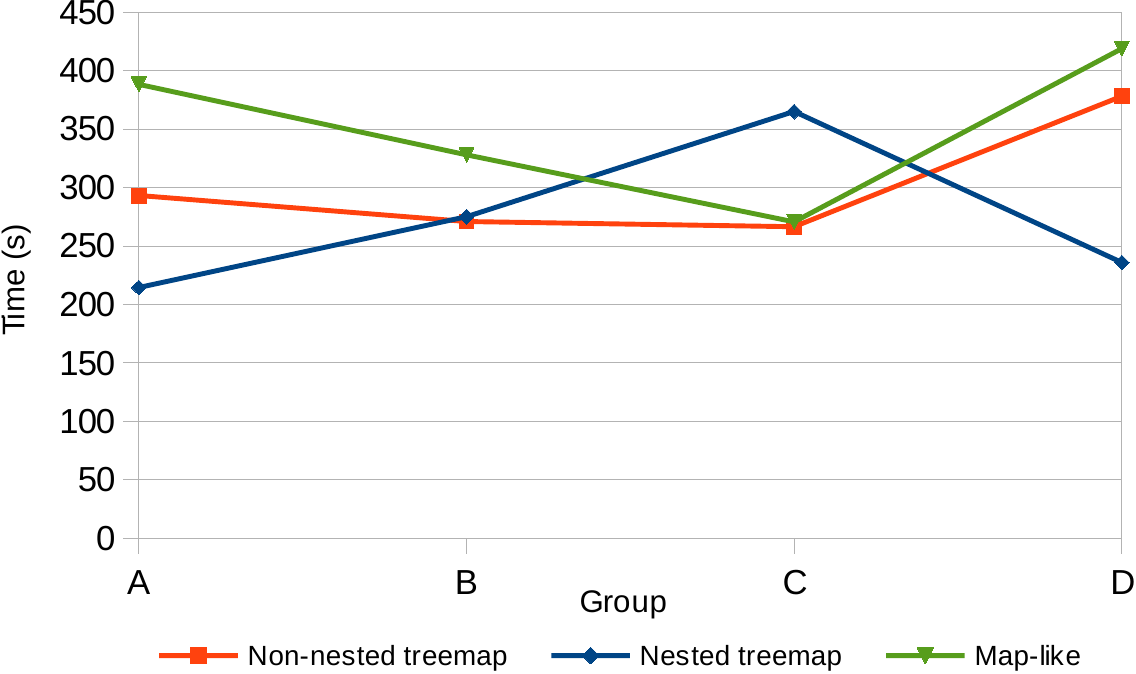}
  \caption{Mean speed of task performance by group}
  \label{fig:speed}
\end{figure}

We measured the time taken to complete each evaluation task. The mean
task completion times per group and per type of visualisation are
summarised in Table~\ref{tab:speed}. Figure~\ref{fig:speed} shows a
plot of these mean task completion times. Counter to our expectation,
we can observe that except in the case of group C, the map-like
visualisation was the slowest type of visualisation to work
with. Treemap was the fastest in all groups. In group B, the
performance times of the two types of treemap were almost
identical. In two of the other groups the nested treemap was clearly
faster than the non-nested treemap, and only in group C was the nested
treemap much slower than the non-nested treemap. Thus it seems that
overall the nested treemap supported the fastest task performance.

\begin{table}
  \centering
  \caption{Mean speed of task performance by group for non-nested
    treemaps, nested treemaps and map-like visualisations (seconds)
    (ranges of the 95\% confidence interval in parentheses)}
  \label{tab:speed}
  \small
  \begin{tabular}{crrrr} \toprule
    \textbf{Group} & \textbf{Non-nested} & \textbf{Nested} & \textbf{Map-like} & \textbf{Avg.} \\ \midrule
    A & 293 ($\pm$ 54.8) & 215 ($\pm$ 26.9) & 388 ($\pm$ 106.3) & 299 \\ 
    B & 271 ($\pm$ 49.9) & 275 ($\pm$ 78.0) & 328 ($\pm$ 75.2) & 291 \\ 
    C & 267 ($\pm$ 76.4) & 365 ($\pm$ 61.9) & 271 ($\pm$ 66.5) & 301 \\ 
    D & 378 ($\pm$ 92.5) & 236 ($\pm$ 50.6) & 419 ($\pm$ 86.6) & 344 \\ \midrule
    Avg. & 302 & 273 & 352 & 309 \\ \bottomrule
  \end{tabular}
\end{table}

\begin{figure}
  \centering
  \includegraphics[width=\columnwidth]{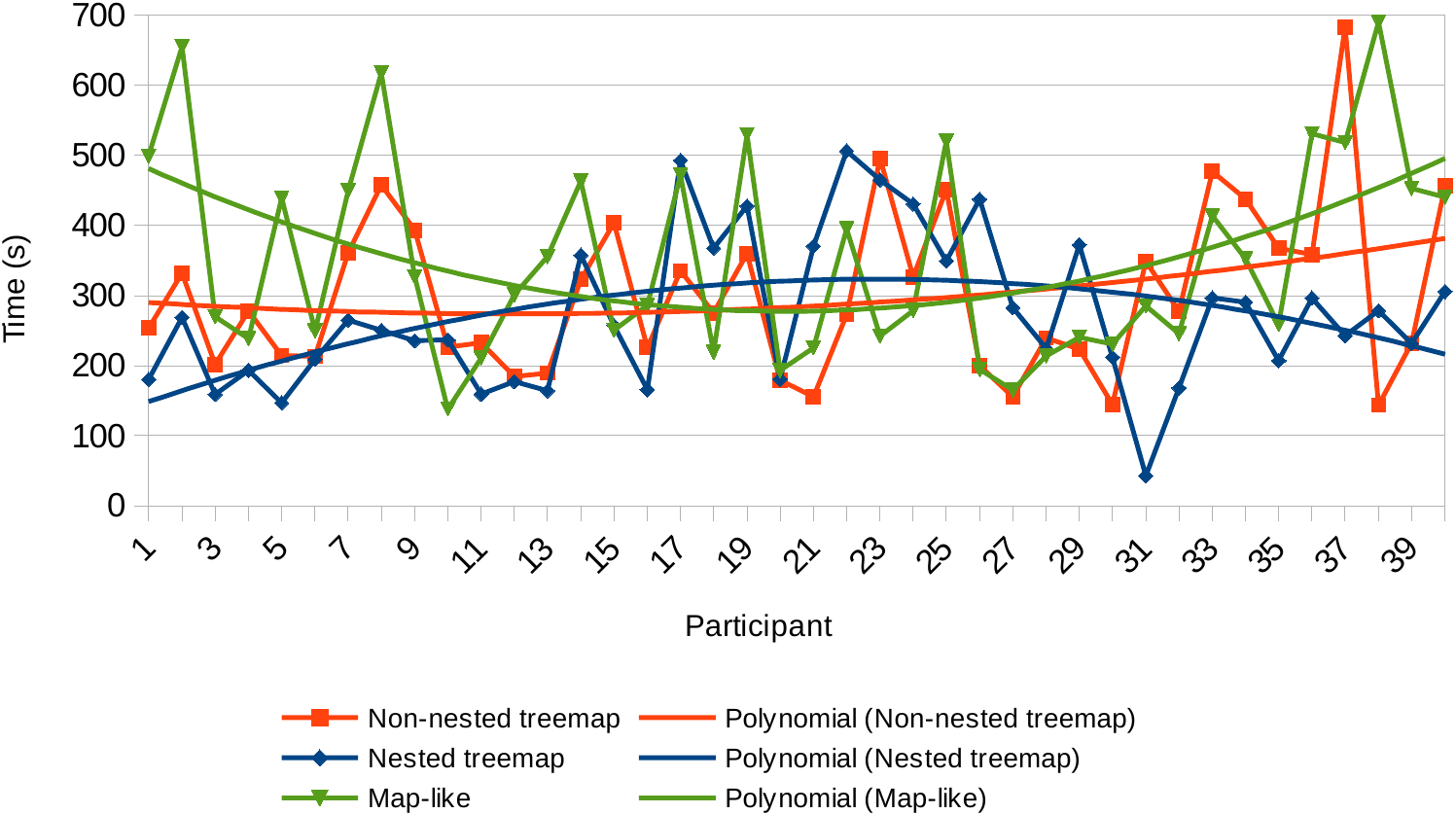}
  \caption{Speed of task performance by individual participants}
  \label{fig:speed-participant}
\end{figure}

To determine the significance of the speed differences we again looked
at individual task performance times for all 40
participants. Figure~\ref{fig:speed-participant} shows the speed
values of each of the 40 participants. Again there is strong
fluctuation across participants, so polynomial trendlines are again
included in the chart.  Detailed speed values of individual
participants for each type of visualisation are shown in
Table~\ref{tab:speed-participant} in
\ref{app:individual-responses}. 

We performed a two-tailed paired t-test and an ANOVA test
on these individual speed figures. As the speed figures of
different types of visualisations strongly overlap we analysed the
significance for all three pairs of visualisations. The results are
shown in Table~\ref{tab:speed-significance}. As the $p$ values show,
the difference in speed between non-nested and nested treemaps was not
statistically significant ($p$ close to 0.2). For the difference
between non-nested treemaps and map-like visualisations the difference
was marginally significant ($p$ close to 0.05). However, for the
difference between nested treemaps and map-like visualisations the
difference was more highly statistically significant ($p <
0.005$). That is, there is statistically significant evidence that
map-like visualisations do \emph{not} allow faster task completion
than treemaps, at least for the type of task we evaluated and
particularly when choosing a nested treemap. Also confirmed by the ANOVA
test ($p < 0.05, F(2, 117) = 4.138, \etaup^2 = 0.066$), the mean speed of
map-like visualisation was in fact higher than those two of treemaps. That
is, map-like visualisations are slower than treemaps. Therefore we 
reject Hypothesis $S_1$ and instead accept the null Hypothesis $S_0$.

\begin{table}
  \centering
  \caption{Statistical significance of speed differences}
  \label{tab:speed-significance}
  \small
  \begin{tabular}{lr} \toprule
    \textbf{Statistical Test} & \textbf{$p$-value} \\ \midrule
    Non-nested vs.\ nested treemap & 0.1955 \\
    Non-nested vs.\ map-like       & 0.0402 \\
    Nested treemap vs.\ map-like   & 0.0048 \\
    ANOVA & 0.0184 \\ \bottomrule
  \end{tabular}
\end{table}

%%%%%%%%%%%%%%%%%%%%%%%%%%%%%%
\subsection{Usability}

This subsection presents the results of the measurements on two usability 
factors, namely the ease of understanding of the data presented in the
visualisations, and the helpfulness of the visualisations for understanding
the data shown in the visualisations.

%%%%%%%%%%%%%%%
\subsubsection{Easiness to Understand}

\begin{figure}
  \centering
  \includegraphics[width=\columnwidth]{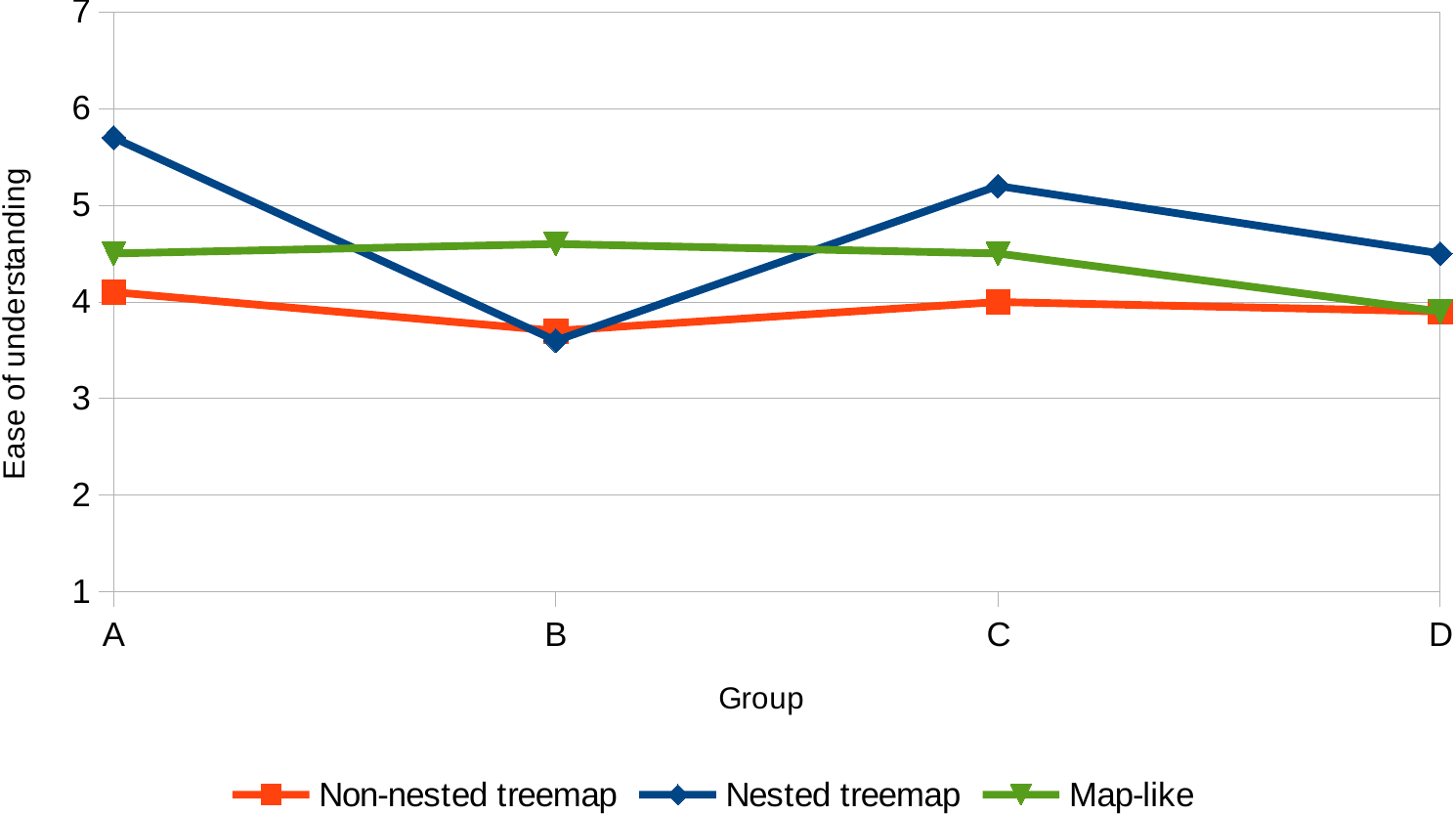}
  \caption{Ease of understanding data through the visualisation, average values by group (values on a Likert scale: 1=very difficult, 7=very easy)}
  \label{fig:ease}
\end{figure}

In the exit questionnaire, we asked the question: ``Do you think it is easy to
understand the data of the University through this visualisation?'' to find 
out the perceived easiness to understand the dataset with each visualisation
used in our experiment. Participants could respond in a 7-point Likert scale
ranging from "very difficult" to "very easy" (where higher scores mean easier).
Table~\ref{tab:ease} and Figure~\ref{fig:ease} shows the results of the easiness
of understanding the data.

\begin{table}
  \centering
  \caption{Ease of understanding data through the visualisation, average values by group (values on a Likert scale: 1=very difficult, 7=very easy)}
  \label{tab:ease}
  \small
  \begin{tabular}{lrrrr} \toprule
    \textbf{Group} & \textbf{Non-nested} & \textbf{Nested} & \textbf{Map-like} & \textbf{Avg.} \\ \midrule
    A & 4.1 & 5.7 & 4.5 & 4.8 \\
    B & 3.7 & 3.6 & 4.6 & 4.0 \\
    C & 4.0 & 5.2 & 4.5 & 4.6 \\
    D & 3.9 & 4.5 & 3.9 & 4.1 \\ \midrule
    Avg. & 3.9 & 4.8 & 4.4 & 4.4 \\ \bottomrule
  \end{tabular}
\end{table}

As shown in Table~\ref{tab:ease}, nested treemaps appeared easiest for users 
to understand the data, followed by the map-like visualisation; meanwhile
non-nested treemaps scored the lowest in this area. However, by inspecting 
Figure~\ref{fig:ease}, we discover that the measurement was fluctuating across
different groups of participants. The ANOVA test further showed an
insignificant result on the comparison among three visualisations ($p = 0.136, 
F(2, 117) = 2.029, \etaup^2 = 0.034$). As such, we cannot conclude that nested treemaps
are better than both non-nested treemaps and map-like visualisations,
in terms of the easiness for understanding the data. However, as 
suggested by the observation of our study, future research can look into 
the factors that make these visualisations perform differently in the
ease-to-understand metric.

%%%%%%%%%%%%%%%
\subsubsection{Helpfulness to Understand}
\begin{figure}
  \centering
  \includegraphics[width=\columnwidth]{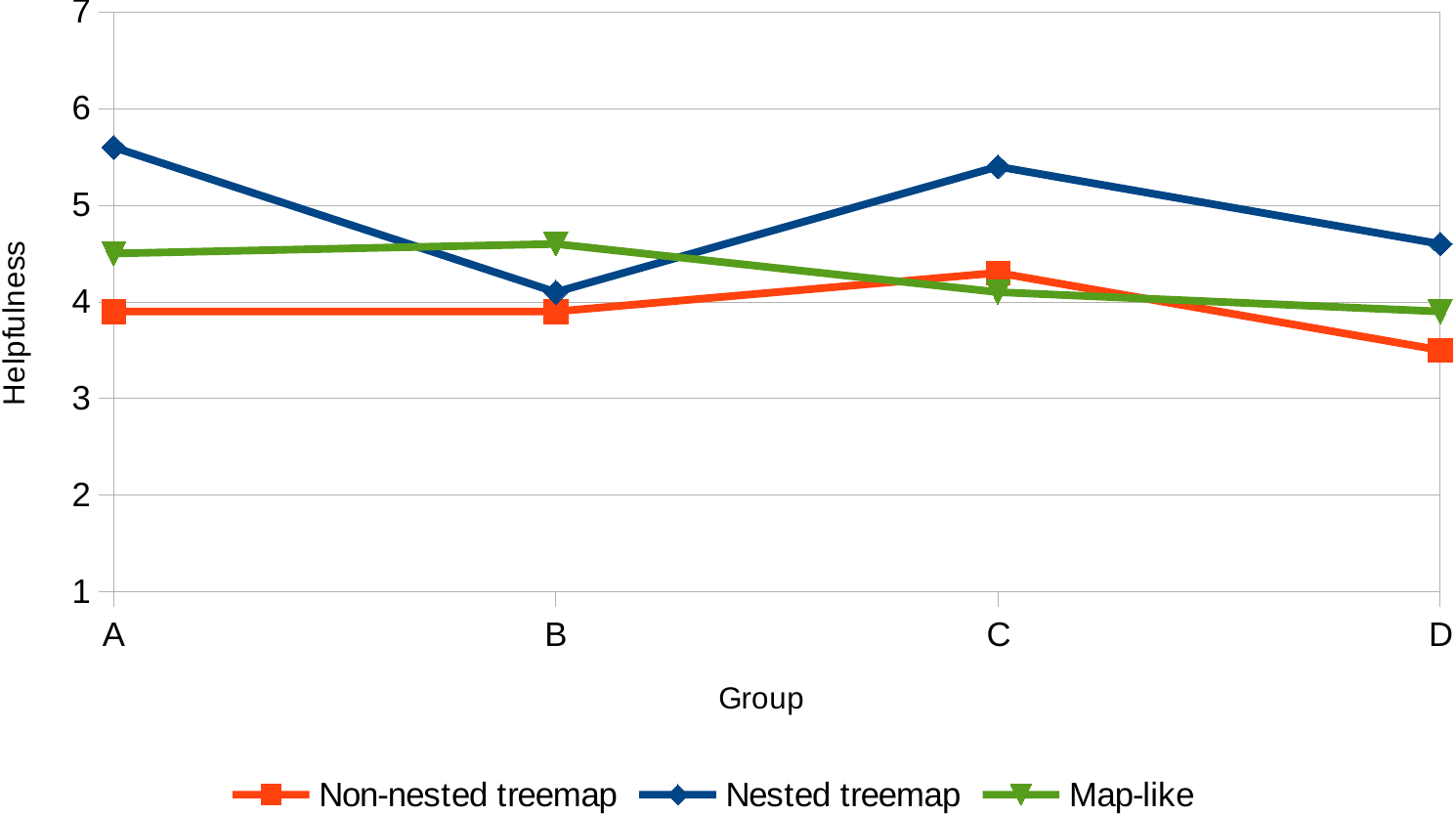}
  \caption{Helpfulness of the visualisation for understanding the data, average values by group (values on a Likert scale: 1=very unhelpful, 7=very helpful)}
  \label{fig:helpfulness}
\end{figure}

Whether a visualisation is perceived helpfulness by the readers is another 
important usability measurement. In order to verify this, we asked another
question: ``Do you think this visualisation helps you to 
understand the data of the University?'' in the exit questionnaire. This helped us to
understand the overall helpfulness of these visualisations. Similarly, participants
required to response to this question with a 7-point Likert scale, ranging from 
"very unhelpful" to "very helpful". A higher score depicted a higher degree of
helpfulness. Table~\ref{tab:helpfulness} and Figure~\ref{fig:helpfulness} list
the results.

\begin{table}
  \centering
  \caption{Helpfulness of the visualisation for understanding the data, average values by group (values on a Likert scale: 1=very unhelpful, 7=very helpful)}
  \label{tab:helpfulness}
  \small
  \begin{tabular}{lrrrr} \toprule
    \textbf{Group} & \textbf{Non-nested} & \textbf{Nested} & \textbf{Map-like} & \textbf{Avg.} \\ \midrule
    A & 3.9 & 5.6 & 4.5 & 4.7 \\
    B & 3.9 & 4.1 & 4.6 & 4.2 \\
    C & 4.3 & 5.4 & 4.1 & 4.6 \\
    D & 3.5 & 4.6 & 3.9 & 4.0 \\ \midrule
    Avg. & 3.9 & 4.9 & 4.3 & 4.4 \\ \bottomrule
  \end{tabular}
\end{table}

Table~\ref{tab:helpfulness} displays the mean values of helpfulness among all three
visualisations. Nested treemaps were perceived the most helpful visualisation for 
understanding the dataset. Map-like visualisation came after nested treemaps, and non-nested
treemaps were worst in our comparison. An ANOVA test showed a significant difference
among all three visualisations ($p < 0.05, F(2, 117) = 3.613, \etaup^2 = 0.058$).
In fact, the mean difference between nested treemaps and map-like visualisations are
small (0.6 out of a 7-point scale). Therefore, we suggest that both nested treemaps and
map-like visualisations are favourable for helping users to understand the data. Both
visualisations are suitable for tasks that allow users to understand the data by
themselves, in which such visualisations serve as a tool for helping users to navigate
and perceive the data.

%%%%%%%%%%%%%%%%%%%%%%%%%%%%%%
\subsection{Qualitative Feedback}

In the final section of the exit questionnaire, we asked the following open-ended
questions for each type of visualisations, in order to capture additional 
qualitative feedback from the participants for their impressions on each 
visualisation:

\begin{quote}
``What do you think are the good or bad points about this type of
visualisation?''
\end{quote}

We received a mix of responses for each visualisation, which we discuss
the main insights with representative quotes below. Grammatical and
spelling mistakes in these quotes are preserved to maintain data integrity.

%%%%%%%%%%%%%%%%%%%%%%%%%%%%%%
\subsubsection{Non-nested Treemap}

The following comments pinpointed the main weakness of non-nested
treemap, which was the difficulty of understanding the hierarchy and the
structure of the underlying data, as compared with nested treemaps.
This was consistent with some prior work about visualising hierarchical 
data with treemaps \cite{Wijk1999, Liang2012, Tu2007}. While
colours and other visual aids might be helpful as suggested by the
comments, the results suggested that non-nested treemaps were not
the best option for visualising hierarchical data. Despite, many 
participants clearly stated that treemaps were good at
demonstrating the size of the individual data item (which is an
university department in our experiment).

\begin{quote}
``It's hard to recognize, which is parent and which is child, because
there's no color or other thing to specific [sic] the relationship.'' (P3)
\end{quote}

\begin{quote}
``In terms of hierarchy, this creates confusion whether the pattern
should be read horizontally or vertically, in or to understand the
relationship from one to another.'' (P4)
\end{quote}

\begin{quote}
``I found it hard to compare two boxes across the image.'' (P9)
\end{quote}

\begin{quote}
``It is not clear to see the relationship between each part of the
department but it clearly shows which department has more student.'' (P25)
\end{quote}

%%%%%%%%%%%%%%%%%%%%%%%%%%%%%%
\subsubsection{Nested Treemap}

Participants agreed that nested treemaps addressed some of the problems found
in non-nested treemaps. Nested treemaps were better for users to recognise the
hierarchy and the structure of the underlying data. However, some participants
commented that nested treemaps were sometimes difficult to understand, and required
specialised knowledge to understand the visualisation. In this regard, 
we reckon that nested treemaps are preferred to non-nested treemap for
showing hierarchical data, but the users' capability for interpreting the
visualisation needs to be considered.

\begin{quote}
``This type of visualization is better than the previous one but again
as it is easier to ascertain the hierarchy and compare the
dimensions. However, it still requires that the user has knowledge of
this type of visualization and can interpret it.'' (P5)
\end{quote}

\begin{quote}
``This is a better pattern compared to the previous one, as it clearly
shows the vertical relationships of the units.'' (P4)
\end{quote}

\begin{quote}
``For the top few lines it's easy to understand; but for some parts at
the bottom it's a bit confused since some lines are not straight
forward which makes me confused about which up-level should it belong
to.'' (P14)
\end{quote}

%%%%%%%%%%%%%%%%%%%%%%%%%%%%%%
\subsection{Map-like Visualisation}

For map-like visualisations, we received a majority of positive comments
and few negative ones. Many participants expressed that it was easy to
clearly understand the hierarchy and different levels of the data. 
In addition, the map metaphor was reassured by the feedback that
the visualisation looked similar to a real geographic map. This further
helps people to understand the underlying information, as discussed
in our prior work \cite{PRMB2017, Pang:2016}.

\begin{quote}
``The good point about this type of visualization is that the map-like
visualization can be much easier to interpret when compared with the
treemap.'' (P15)
\end{quote}

\begin{quote}
``The map-like visualization is by far (of all the presented) the
easiest one to interpret as everything just seems (or appears) much
simpler. One can easily determine which variables are at the same or
different level.'' (P5)
\end{quote}

\begin{quote}
``It looks like a geographical map.'' (P4)
\end{quote}

\begin{quote}
``... similar to maps makes people more familiar with it.'' (P30)
\end{quote}

\begin{quote}
``To me, the information is easily read. And it more direct than a pie
chart and more vivid than a Treemap.'' (P21)
\end{quote}

On the other hand, a minority of comments reported that the visualisation 
looked strange because it did not use the conventional visual 
representation (such as rectangles and circles) than other information
visualisations, which made it hard to compare the size of different data
items. Also, the positions of the text labels were crucial for
describing the data hierarchy, and glitches in the visualisation software
caused them to be misplaced. This needs to be addressed in future 
implementations to avoid misinterpretations of the data.

\begin{quote}
``It looks a bit strange as the irregular shape is very different from
the conventional square or circle chart we use.'' (P22)
\end{quote}

\begin{quote}
``Furthermore, these types of visualizations don't use numbers
or percentages so there is no need to actually worry about comparing
those, one simply has to look at the different visual
dimensions. However, this can also prove to be a bad point as there is
no way of determining the numbers, one may only infer.'' (P5)
\end{quote}

\begin{quote}
``The position of the name have to be precise, otherwise it may be
troublesome to identify the relationship between elements/sectors.'' (P35)
\end{quote}

Overall, according to the feedback, the map-like visualisation shows
a promising result of representing hierarchical data in an easily
readable and understandable manner. It has a potential to be used as an
alternative of both types of treemaps.

%%%%%%%%%%%%%%%%%%%%%%%%%%%%%%%%%%%%%%%%%%%%%%%%%%%%%%%%%%%%
\section{Conclusions}

Over the past decade map-like visualisation has emerged as an
attractive tool in the toolbox of the visualisation practitioner, one
that has the advantage of ready understandability without prior
training thanks to the wide exposure to maps in society. This makes it
desirable to understand the strengths and potential weaknesses of this
type of visualisation. In this paper we presented an evaluation that
sought to test whether or not our assumption that map-like
visualisations are more accurate and faster to use than treemaps is
true.

The results and statistical analysis of our evaluation lead us to two
main conclusions: (1) map-like visualisations are indeed better in terms of
accuracy than both forms of treemaps we evaluated; (2) of the two
types of treemap evaluated, the non-nested treemap performs very
poorly in terms of accuracy. Our results therefore suggest that for
tasks requiring the accurate recognition of hierarchy, a map-like
visualisation should be preferred over a treemap; and that if a
treemap must be employed for such tasks then the nested treemap should
be greatly preferred over the non-nested treemap.

However, our results also showed that, at least in the experiment
setup of our evaluation, map-like visualisations are slower than
treemaps, even significantly slower in the case of comparing against
nested treemaps. Nonetheless, in cases where speed is less important
than accuracy, the use of map-like visualisations would still be
beneficial. We note that the slower speed of the use of map-like
visualisations warrants further study, perhaps exploring other types
of map-like visualisation and other tasks.

Finally, in terms of statistical analysis of the usability factors, 
we cannot judge that any visualisation is easier for understanding
the data than the others. However, nested treemaps are slightly more
helpful than map-like visualisations for understanding the data, while
non-nested treemaps perform very poor in this regard. Combining with the
analysis of the qualitative feedback, we conclude that map-like 
visualisations is promising to be used as an alternative of treemaps,
particularly where accuracy is required.

%%%%%%%%%%%%%%%%%%%%%%%%%%%%%%%%%%%%%%%%%%%%%%%%%%%%%%%%%%%%
\section{Acknowledgements}

The support by the University of Macau Research Committee under grant
number MYRG2014-00172-FST is gratefully acknowledged. We also 
acknowledge the help from Bin Pang for facilitating the experiment.

\section{References}
\bibliographystyle{elsarticle-num}
\bibliography{JVLC2018}

\appendix

\section{Individual Survey Responses}
\label{app:individual-responses}

The following tables show the detailed accuracy and speed values of
all participants of our survey.

\begin{table}
  \centering
  \caption{Accuracy for non-nested treemaps, nested treemaps and
    map-like visualisations by individual participants}
  \label{tab:acc-participant}
  \small
  \begin{tabular}{rrrr} \toprule
    \textbf{Participant} & \textbf{Non-nested} & \textbf{Nested} & \textbf{Map-like} \\ \midrule
    1 & 10\% & 70\% & 80\% \\ 
    2 & 10\% & 50\% & 100\% \\ 
    3 & 70\% & 100\% & 90\% \\ 
    4 & 20\% & 60\% & 90\% \\ 
    5 & 20\% & 80\% & 100\% \\ 
    6 & 40\% & 70\% & 80\% \\ 
    7 & 0\% & 90\% & 90\% \\ 
    8 & 0\% & 90\% & 100\% \\ 
    9 & 20\% & 50\% & 90\% \\ 
    10 & 20\% & 70\% & 80\% \\ 
    11 & 60\% & 40\% & 30\% \\ 
    12 & 40\% & 70\% & 90\% \\ 
    13 & 60\% & 100\% & 80\% \\ 
    14 & 50\% & 90\% & 80\% \\ 
    15 & 30\% & 80\% & 80\% \\ 
    16 & 20\% & 40\% & 40\% \\ 
    17 & 60\% & 80\% & 80\% \\ 
    18 & 10\% & 100\% & 100\% \\ 
    19 & 20\% & 70\% & 90\% \\ 
    20 & 30\% & 70\% & 80\% \\ 
    21 & 30\% & 90\% & 100\% \\ 
    22 & 20\% & 50\% & 70\% \\ 
    23 & 20\% & 40\% & 70\% \\ 
    24 & 40\% & 90\% & 100\% \\ 
    25 & 70\% & 80\% & 100\% \\ 
    26 & 30\% & 80\% & 100\% \\ 
    27 & 50\% & 100\% & 80\% \\ 
    28 & 20\% & 70\% & 100\% \\ 
    29 & 30\% & 80\% & 100\% \\ 
    30 & 10\% & 60\% & 20\% \\ 
    31 & 0\% & 0\% & 100\% \\ 
    32 & 30\% & 70\% & 80\% \\ 
    33 & 30\% & 100\% & 80\% \\ 
    34 & 30\% & 90\% & 100\% \\ 
    35 & 30\% & 90\% & 100\% \\ 
    36 & 20\% & 80\% & 100\% \\ 
    37 & 60\% & 70\% & 100\% \\ 
    38 & 10\% & 50\% & 100\% \\ 
    39 & 40\% & 60\% & 70\% \\ 
    40 & 30\% & 40\% & 100\% \\ \bottomrule
  \end{tabular}
\end{table}

\begin{table}
  \centering
  \caption{Speed of task performance for non-nested treemaps, nested
    treemaps and map-like visualisations by individual participants
    (seconds)}
  \label{tab:speed-participant}
  \small
  \begin{tabular}{rrrr} \toprule
    \textbf{ID} & \textbf{Non-nested treemap} & \textbf{Nested treemap} & \textbf{Map-like} \\ \midrule
    1 & 254 & 180 & 499 \\ 
    2 & 332 & 269 & 655 \\ 
    3 & 202 & 159 & 270 \\ 
    4 & 278 & 194 & 239 \\ 
    5 & 215 & 147 & 439 \\ 
    6 & 213 & 209 & 250 \\ 
    7 & 361 & 265 & 449 \\ 
    8 & 458 & 251 & 618 \\ 
    9 & 393 & 236 & 327 \\ 
    10 & 227 & 237 & 138 \\ 
    11 & 233 & 159 & 211 \\ 
    12 & 185 & 177 & 301 \\ 
    13 & 190 & 164 & 356 \\ 
    14 & 323 & 358 & 463 \\ 
    15 & 404 & 257 & 251 \\ 
    16 & 226 & 166 & 286 \\ 
    17 & 335 & 493 & 472 \\ 
    18 & 276 & 368 & 219 \\ 
    19 & 360 & 428 & 530 \\ 
    20 & 179 & 181 & 193 \\ 
    21 & 156 & 370 & 226 \\ 
    22 & 273 & 507 & 395 \\ 
    23 & 496 & 465 & 242 \\ 
    24 & 327 & 431 & 279 \\ 
    25 & 451 & 349 & 521 \\ 
    26 & 200 & 437 & 195 \\ 
    27 & 156 & 283 & 165 \\ 
    28 & 240 & 225 & 214 \\ 
    29 & 224 & 373 & 241 \\ 
    30 & 144 & 212 & 231 \\ 
    31 & 349 & 43 & 285 \\ 
    32 & 277 & 168 & 246 \\ 
    33 & 477 & 297 & 414 \\ 
    34 & 438 & 290 & 353 \\ 
    35 & 368 & 207 & 259 \\ 
    36 & 359 & 297 & 531 \\ 
    37 & 683 & 243 & 519 \\ 
    38 & 143 & 279 & 690 \\ 
    39 & 232 & 232 & 453 \\ 
    40 & 456 & 306 & 440 \\ \bottomrule
  \end{tabular}
\end{table}

\begin{table}
  \centering
  \caption{Perceived ease-of-use for non-nested treemaps, nested
    treemaps and map-like visualisations by individual participants
    (7-point Likert scale)}
  \label{tab:easeofuse-participant}
  \small
  \begin{tabular}{rrrr} \toprule
    \textbf{ID} & \textbf{Non-nested treemap} & \textbf{Nested treemap} & \textbf{Map-like} \\ \midrule
    1  & 1 & 5 & 1 \\
    2  & 5 & 7 & 7 \\
    3  & 1 & 6 & 4 \\
    4  & 6 & 5 & 2 \\
    5  & 2 & 3 & 5 \\
    6  & 6 & 6 & 5 \\
    7  & 4 & 6 & 5 \\
    8  & 6 & 6 & 6 \\
    9  & 5 & 6 & 4 \\
    10 & 5 & 7 & 6 \\
    11 & 1 & 4 & 5 \\
    12 & 5 & 6 & 4 \\
    13 & 6 & 5 & 3 \\
    14 & 2 & 2 & 7 \\
    15 & 5 & 2 & 7 \\
    16 & 2 & 2 & 2 \\
    17 & 7 & 1 & 6 \\
    18 & 4 & 7 & 7 \\
    19 & 2 & 4 & 1 \\
    20 & 3 & 3 & 4 \\
    21 & 3 & 4 & 5 \\
    22 & 7 & 7 & 2 \\
    23 & 5 & 5 & 4 \\
    24 & 2 & 6 & 4 \\
    25 & 2 & 5 & 3 \\
    26 & 1 & 4 & 3 \\
    27 & 2 & 6 & 6 \\
    28 & 7 & 2 & 7 \\
    29 & 6 & 6 & 7 \\
    30 & 5 & 7 & 4 \\
    31 & 5 & 4 & 5 \\
    32 & 3 & 2 & 1 \\
    33 & 5 & 6 & 3 \\
    34 & 5 & 5 & 3 \\
    35 & 4 & 3 & 5 \\
    36 & 4 & 4 & 4 \\
    37 & 6 & 7 & 6 \\
    38 & 1 & 3 & 6 \\
    39 & 1 & 7 & 4 \\
    40 & 5 & 4 & 2 \\ \bottomrule
  \end{tabular}
\end{table}

\begin{table}
  \centering
  \caption{Perceived helpfulness for non-nested treemaps, nested
    treemaps and map-like visualisations by individual participants
    (7-point Likert scale)}
  \label{tab:helpfulness-participant}
  \small
  \begin{tabular}{rrrr} \toprule
    \textbf{ID} & \textbf{Non-nested treemap} & \textbf{Nested treemap} & \textbf{Map-like} \\ \midrule
    1  & 1 & 5 & 1 \\
    2  & 6 & 7 & 7 \\
    3  & 1 & 6 & 4 \\
    4  & 5 & 6 & 3 \\
    5  & 2 & 3 & 5 \\
    6  & 7 & 6 & 5 \\
    7  & 4 & 5 & 5 \\
    8  & 5 & 6 & 6 \\
    9  & 5 & 5 & 3 \\
    10 & 3 & 7 & 6 \\
    11 & 2 & 4 & 5 \\
    12 & 5 & 5 & 4 \\
    13 & 4 & 5 & 3 \\
    14 & 1 & 5 & 7 \\
    15 & 6 & 3 & 6 \\
    16 & 5 & 5 & 3 \\
    17 & 7 & 1 & 6 \\
    18 & 4 & 6 & 7 \\
    19 & 2 & 4 & 1 \\
    20 & 3 & 3 & 4 \\
    21 & 4 & 5 & 6 \\
    22 & 7 & 7 & 2 \\
    23 & 5 & 4 & 4 \\
    24 & 3 & 6 & 4 \\
    25 & 2 & 6 & 2 \\
    26 & 2 & 4 & 3 \\
    27 & 2 & 6 & 6 \\
    28 & 7 & 3 & 5 \\
    29 & 6 & 6 & 4 \\
    30 & 5 & 7 & 5 \\
    31 & 3 & 4 & 4 \\
    32 & 4 & 4 & 1 \\
    33 & 5 & 7 & 4 \\
    34 & 4 & 5 & 3 \\
    35 & 4 & 4 & 7 \\
    36 & 1 & 1 & 2 \\
    37 & 6 & 6 & 6 \\
    38 & 2 & 4 & 6 \\
    39 & 1 & 7 & 4 \\
    40 & 5 & 4 & 2 \\ \bottomrule
  \end{tabular}
\end{table}

\end{document}

%% file: figures/treemap-sample.tex
\setlength{\unitlength}{4144sp}%
\begingroup\makeatletter\ifx\SetFigFont\undefined%
\gdef\SetFigFont#1#2#3#4#5{%
  \reset@font\fontsize{#1}{#2pt}%
  \fontfamily{#3}\fontseries{#4}\fontshape{#5}%
  \selectfont}%
\fi\endgroup%
\begin{picture}(1734,2637)(439,-2233)
\thinlines
{\color[rgb]{0,0,0}\put(541,-781){\framebox(1530,990){}}
}%
{\color[rgb]{0,0,0}\put(1171,-1861){\framebox(270,180){}}
}%
{\color[rgb]{0,0,0}\put(1531,-1861){\framebox(270,180){}}
}%
{\color[rgb]{0,0,0}\put(1081,-1951){\framebox(810,450){}}
}%
{\color[rgb]{0,0,0}\put(541,-2131){\framebox(1530,1260){}}
}%
{\color[rgb]{0,0,0}\put(451,-2221){\framebox(1710,2610){}}
}%
{\color[rgb]{0,0,0}\put(1711,-691){\framebox(270,720){}}
}%
{\color[rgb]{0,0,0}\put(631,-691){\framebox(990,720){}}
}%
{\color[rgb]{0,0,0}\put(721,-331){\framebox(810,180){}}
}%
{\color[rgb]{0,0,0}\put(721,-601){\framebox(810,180){}}
}%
{\color[rgb]{0,0,0}\put(631,-2041){\framebox(270,990){}}
}%
{\color[rgb]{0,0,0}\put(991,-2041){\framebox(990,990){}}
}%
{\color[rgb]{0,0,0}\put(1081,-1411){\framebox(810,180){}}
}%
\put(541,254){\makebox(0,0)[b]{\smash{{\SetFigFont{8}{9.6}{\sfdefault}{\mddefault}{\updefault}{\color[rgb]{0,0,0}A}%
}}}}
\put(631, 74){\makebox(0,0)[b]{\smash{{\SetFigFont{8}{9.6}{\sfdefault}{\mddefault}{\updefault}{\color[rgb]{0,0,0}B}%
}}}}
\put(721,-106){\makebox(0,0)[b]{\smash{{\SetFigFont{8}{9.6}{\sfdefault}{\mddefault}{\updefault}{\color[rgb]{0,0,0}C}%
}}}}
\put(811,-286){\makebox(0,0)[b]{\smash{{\SetFigFont{8}{9.6}{\sfdefault}{\mddefault}{\updefault}{\color[rgb]{0,0,0}E}%
}}}}
\put(631,-1006){\makebox(0,0)[b]{\smash{{\SetFigFont{8}{9.6}{\sfdefault}{\mddefault}{\updefault}{\color[rgb]{0,0,0}G}%
}}}}
\put(721,-1186){\makebox(0,0)[b]{\smash{{\SetFigFont{8}{9.6}{\sfdefault}{\mddefault}{\updefault}{\color[rgb]{0,0,0}H}%
}}}}
\put(1261,-1816){\makebox(0,0)[b]{\smash{{\SetFigFont{8}{9.6}{\sfdefault}{\mddefault}{\updefault}{\color[rgb]{0,0,0}L}%
}}}}
\put(1621,-1816){\makebox(0,0)[b]{\smash{{\SetFigFont{8}{9.6}{\sfdefault}{\mddefault}{\updefault}{\color[rgb]{0,0,0}M}%
}}}}
\put(1171,-1636){\makebox(0,0)[b]{\smash{{\SetFigFont{8}{9.6}{\sfdefault}{\mddefault}{\updefault}{\color[rgb]{0,0,0}K}%
}}}}
\put(1801,-106){\makebox(0,0)[b]{\smash{{\SetFigFont{8}{9.6}{\sfdefault}{\mddefault}{\updefault}{\color[rgb]{0,0,0}D}%
}}}}
\put(811,-556){\makebox(0,0)[b]{\smash{{\SetFigFont{8}{9.6}{\sfdefault}{\mddefault}{\updefault}{\color[rgb]{0,0,0}F}%
}}}}
\put(1081,-1186){\makebox(0,0)[b]{\smash{{\SetFigFont{8}{9.6}{\sfdefault}{\mddefault}{\updefault}{\color[rgb]{0,0,0}I}%
}}}}
\put(1171,-1366){\makebox(0,0)[b]{\smash{{\SetFigFont{8}{9.6}{\sfdefault}{\mddefault}{\updefault}{\color[rgb]{0,0,0}J}%
}}}}
\end{picture}%

%% file: JVLC2018.bbl
\begin{thebibliography}{10}
\expandafter\ifx\csname url\endcsname\relax
  \def\url#1{\texttt{#1}}\fi
\expandafter\ifx\csname urlprefix\endcsname\relax\def\urlprefix{URL }\fi
\expandafter\ifx\csname href\endcsname\relax
  \def\href#1#2{#2} \def\path#1{#1}\fi

\bibitem{Hall2013}
M.~Hall, R.~M. Kirby, F.~Li, M.~Meyer, V.~Pascucci, J.~M. Phillips, R.~Ricci,
  J.~Van~der Merwe, S.~Venkatasubramanian, Rethinking abstractions for big
  data: Why, where, how, and what, arXiv preprint arXiv:1306.3295.

\bibitem{Wills2009}
G.~Wills,
  \href{http://dx.doi.org/10.1007/978-0-387-39940-9{\_}1380}{Visualizing
  hierarchical data}, in: L.~Liu, M.~T. {\"{O}}zsu (Eds.), Encyclopedia of
  Database Systems, Springer US, Boston, MA, USA, 2009, pp. 3425--3432.
\newblock \href {http://dx.doi.org/10.1007/978-0-387-39940-9_1380}
  {\path{doi:10.1007/978-0-387-39940-9_1380}}.
\newline\urlprefix\url{http://dx.doi.org/10.1007/978-0-387-39940-9{\_}1380}

\bibitem{JS:1991}
B.~Johnson, B.~Shneiderman, Tree-maps: A space-filling approach to the
  visualization of hierarchical information structures, in: Proceedings of the
  IEEE Conference on Visualization, IEEE, Piscataway, NJ, USA, 1991, pp.
  284--291.
\newblock \href {http://dx.doi.org/10.1109/VISUAL.1991.175815}
  {\path{doi:10.1109/VISUAL.1991.175815}}.

\bibitem{Shn:1992}
B.~Shneiderman, \href{http://doi.acm.org/10.1145/102377.115768}{Tree
  visualization with tree-maps: 2-{D} space-filling approach}, ACM Transactions
  on Graphics 11~(1) (1992) 92--99.
\newblock \href {http://dx.doi.org/10.1145/102377.115768}
  {\path{doi:10.1145/102377.115768}}.
\newline\urlprefix\url{http://doi.acm.org/10.1145/102377.115768}

\bibitem{Bruls2000}
M.~Bruls, K.~Huizing, J.~J. van Wijk,
  \href{http://dx.doi.org/10.1007/978-3-7091-6783-0{\_}4}{Squarified treemaps},
  in: W.~C. de~Leeuw, R.~van Liere (Eds.), Data Visualization 2000: Proceedings
  of the Joint EUROGRAPHICS and IEEE TCVG Symposium on Visualization, Springer
  Vienna, Vienna, Austria, 2000, pp. 33--42.
\newblock \href {http://dx.doi.org/10.1007/978-3-7091-6783-0_4}
  {\path{doi:10.1007/978-3-7091-6783-0_4}}.
\newline\urlprefix\url{http://dx.doi.org/10.1007/978-3-7091-6783-0{\_}4}

\bibitem{Wijk1999}
J.~J. {van Wijk}, H.~{van de Wetering}, Cushion treemaps: visualization of
  hierarchical information, in: Proceedings of the 1999 IEEE Symposium on
  Information Visualization, IEEE Computer Society, Los Alamitos, CA, USA,
  1999, pp. 73--78,147.
\newblock \href {http://dx.doi.org/10.1109/INFVIS.1999.801860}
  {\path{doi:10.1109/INFVIS.1999.801860}}.

\bibitem{Schreck2006}
T.~Schreck, D.~Keim, F.~Mansmann,
  \href{http://doi.acm.org/10.1145/2602161.2602183}{Regular treemap layouts for
  visual analysis of hierarchical data}, in: Proceedings of the 22nd Spring
  Conference on Computer Graphics, SCCG '06, ACM, New York, NY, USA, 2006, pp.
  183--190.
\newblock \href {http://dx.doi.org/10.1145/2602161.2602183}
  {\path{doi:10.1145/2602161.2602183}}.
\newline\urlprefix\url{http://doi.acm.org/10.1145/2602161.2602183}

\bibitem{Liang2012}
J.~Liang, Q.~V. Nguyen, S.~Simoff, M.~L. Huang, Angular treemaps --- a new
  technique for visualizing and emphasizing hierarchical structures, in: 16th
  International Conference on Information Visualisation, IEEE Computer Society,
  Los Alamitos, CA, USA, 2012, pp. 74--80.
\newblock \href {http://dx.doi.org/10.1109/IV.2012.23}
  {\path{doi:10.1109/IV.2012.23}}.

\bibitem{Tu2007}
Y.~Tu, H.-W. Shen, Visualizing changes of hierarchical data using treemaps,
  IEEE Transactions on Visualization and Computer Graphics 13~(6) (2007)
  1286--1293.
\newblock \href {http://dx.doi.org/10.1109/TVCG.2007.70529}
  {\path{doi:10.1109/TVCG.2007.70529}}.

\bibitem{BAY2015}
R.~P. Biuk-Aghai, M.~Yang, P.~C.-I. Pang, W.~H. Ao, S.~Fong, Y.-W. Si,
  \href{http://www.sciencedirect.com/science/article/pii/S1045926X15000609}{A
  map-like visualisation method based on liquid modelling}, Journal of Visual
  Languages {\&} Computing 31, Part A (2015) 87--103.
\newblock \href
  {http://dx.doi.org/http://dx.doi.org/10.1016/j.jvlc.2015.10.003}
  {\path{doi:http://dx.doi.org/10.1016/j.jvlc.2015.10.003}}.
\newline\urlprefix\url{http://www.sciencedirect.com/science/article/pii/S1045926X15000609}

\bibitem{YBA:2015}
M.~Yang, R.~P. Biuk-Aghai,
  \href{http://doi.acm.org/10.1145/2801040.2801056}{Enhanced hexagon-tiling
  algorithm for map-like information visualisation}, in: Proceedings of the 8th
  International Symposium on Visual Information Communication and Interaction,
  VINCI '15, ACM, New York, NY, USA, 2015, pp. 137--142.
\newblock \href {http://dx.doi.org/10.1145/2801040.2801056}
  {\path{doi:10.1145/2801040.2801056}}.
\newline\urlprefix\url{http://doi.acm.org/10.1145/2801040.2801056}

\bibitem{Pang:2011}
C.-I. Pang, R.~P. Biuk-Aghai,
  \href{http://doi.acm.org/10.1145/2038558.2038579}{Wikipedia world map: Method
  and application of map-like wiki visualization}, in: Proceedings of the 7th
  International Symposium on Wikis and Open Collaboration, WikiSym '11, ACM,
  New York, NY, USA, 2011, pp. 124--133.
\newblock \href {http://dx.doi.org/10.1145/2038558.2038579}
  {\path{doi:10.1145/2038558.2038579}}.
\newline\urlprefix\url{http://doi.acm.org/10.1145/2038558.2038579}

\bibitem{BAPS2014}
R.~P. Biuk-Aghai, C.-I. Pang, Y.-W. Si,
  \href{http://www.sciencedirect.com/science/article/pii/S0167739X13000617}{Visualizing
  large-scale human collaboration in wikipedia}, Future Generation Computer
  Systems 31 (2014) 120--133.
\newblock \href
  {http://dx.doi.org/http://dx.doi.org/10.1016/j.future.2013.04.001}
  {\path{doi:http://dx.doi.org/10.1016/j.future.2013.04.001}}.
\newline\urlprefix\url{http://www.sciencedirect.com/science/article/pii/S0167739X13000617}

\bibitem{PBY2016}
P.~C.-I. Pang, R.~P. Biuk-Aghai, M.~Yang,
  \href{http://doi.acm.org/10.1145/2968220.2968239}{What makes you think this
  is a map?: Suggestions for creating map-like visualisations}, in: Proceedings
  of the 9th International Symposium on Visual Information Communication and
  Interaction, VINCI '16, ACM, New York, NY, USA, 2016, pp. 75--82.
\newblock \href {http://dx.doi.org/10.1145/2968220.2968239}
  {\path{doi:10.1145/2968220.2968239}}.
\newline\urlprefix\url{http://doi.acm.org/10.1145/2968220.2968239}

\bibitem{BPP2017}
R.~P. Biuk-Aghai, P.~C.-I. Pang, B.~Pang,
  \href{http://doi.acm.org/10.1145/3105971.3105976}{Map-like visualisations vs.
  treemaps: An experimental comparison}, in: Proceedings of the 10th
  International Symposium on Visual Information Communication and Interaction,
  VINCI '17, ACM, New York, NY, USA, 2017, pp. 113--120.
\newblock \href {http://dx.doi.org/10.1145/3105971.3105976}
  {\path{doi:10.1145/3105971.3105976}}.
\newline\urlprefix\url{http://doi.acm.org/10.1145/3105971.3105976}

\bibitem{Bederson2002}
B.~B. Bederson, B.~Shneiderman, M.~Wattenberg,
  \href{http://doi.acm.org/10.1145/571647.571649}{Ordered and quantum treemaps:
  Making effective use of {2D} space to display hierarchies}, ACM Trans. Graph.
  21~(4) (2002) 833--854.
\newblock \href {http://dx.doi.org/10.1145/571647.571649}
  {\path{doi:10.1145/571647.571649}}.
\newline\urlprefix\url{http://doi.acm.org/10.1145/571647.571649}

\bibitem{Kosara}
R.~Kosara, \href{https://eagereyes.org/techniques/treemaps}{Treemaps} (2008).
\newline\urlprefix\url{https://eagereyes.org/techniques/treemaps}

\bibitem{LNSH2015}
J.~Liang, Q.~V. Nguyen, S.~Simoff, M.~L. Huang,
  \href{http://www.sciencedirect.com/science/article/pii/S1045926X1500066X}{Divide
  and conquer treemaps: Visualizing large trees with various shapes}, Journal
  of Visual Languages {\&} Computing 31 (2015) 104 -- 127.
\newblock \href
  {http://dx.doi.org/http://dx.doi.org/10.1016/j.jvlc.2015.10.009}
  {\path{doi:http://dx.doi.org/10.1016/j.jvlc.2015.10.009}}.
\newline\urlprefix\url{http://www.sciencedirect.com/science/article/pii/S1045926X1500066X}

\bibitem{Hahn2015}
S.~Hahn, Comparing the layout stability of treemap algorithms, Proceedings of
  the 8th Ph.D. Retreat of the HPI Research School on Service-oriented Systems
  Engineering 95 (2015) 71.

\bibitem{deBerg2014}
M.~de~Berg, B.~Speckmann, V.~van~der Weele,
  \href{http://www.sciencedirect.com/science/article/pii/S0925772113001764}{Treemaps
  with bounded aspect ratio}, Computational Geometry 47~(6) (2014) 683 -- 693,
  27th European Workshop on Computational Geometry (EuroCG 2011).
\newblock \href
  {http://dx.doi.org/https://doi.org/10.1016/j.comgeo.2013.12.008}
  {\path{doi:https://doi.org/10.1016/j.comgeo.2013.12.008}}.
\newline\urlprefix\url{http://www.sciencedirect.com/science/article/pii/S0925772113001764}

\bibitem{Hahn2014}
S.~Hahn, J.~Tr{\"u}mper, D.~Moritz, J.~D{\"o}llner, Visualization of varying
  hierarchies by stable layout of voronoi treemaps, in: 2014 International
  Conference on Information Visualization Theory and Applications (IVAPP),
  IEEE, 2014, pp. 50--58.

\bibitem{CP:2012}
A.~Celentano, F.~Pittarello,
  \href{http://dx.doi.org/10.1016/j.jvlc.2011.11.004}{From real to metaphoric
  maps: Cartography as a visual language for organizing and sharing knowledge},
  Journal of Visual Languages and Computing 23~(2) (2012) 63--77.
\newblock \href {http://dx.doi.org/10.1016/j.jvlc.2011.11.004}
  {\path{doi:10.1016/j.jvlc.2011.11.004}}.
\newline\urlprefix\url{http://dx.doi.org/10.1016/j.jvlc.2011.11.004}

\bibitem{Couclelis1998}
H.~Couclelis,
  \href{http://www.tandfonline.com/doi/abs/10.1559/152304098782383034}{Worlds
  of information: The geographic metaphor in the visualization of complex
  information}, Cartography and Geographic Information Systems 25~(4) (1998)
  209--220.
\newblock \href {http://dx.doi.org/10.1559/152304098782383034}
  {\path{doi:10.1559/152304098782383034}}.
\newline\urlprefix\url{http://www.tandfonline.com/doi/abs/10.1559/152304098782383034}

\bibitem{Borner2003}
K.~B{\"{o}}rner, C.~Chen, K.~W. Boyack,
  \href{http://dx.doi.org/10.1002/aris.1440370106}{Visualizing knowledge
  domains}, Annual Review of Information Science and Technology 37~(1) (2003)
  179--255.
\newblock \href {http://dx.doi.org/10.1002/aris.1440370106}
  {\path{doi:10.1002/aris.1440370106}}.
\newline\urlprefix\url{http://dx.doi.org/10.1002/aris.1440370106}

\bibitem{Borner2010}
K.~B{\"{o}}rner, Atlas of Science, MIT Press, Cambridge, MA, USA, 2010.

\bibitem{Skupin2004}
A.~Skupin, \href{http://www.pnas.org/content/101/suppl{\_}1/5274.abstract}{The
  world of geography: Visualizing a knowledge domain with cartographic means},
  Proceedings of the National Academy of Sciences 101~(suppl 1) (2004)
  5274--5278.
\newblock \href {http://dx.doi.org/10.1073/pnas.0307654100}
  {\path{doi:10.1073/pnas.0307654100}}.
\newline\urlprefix\url{http://www.pnas.org/content/101/suppl{\_}1/5274.abstract}

\bibitem{Mashima2012}
D.~Mashima, S.~Kobourov, Y.~Hu, Visualizing dynamic data with maps, IEEE
  Transactions on Visualization and Computer Graphics 18~(9) (2012) 1424--1437.
\newblock \href {http://dx.doi.org/10.1109/TVCG.2011.288}
  {\path{doi:10.1109/TVCG.2011.288}}.

\bibitem{Gronemann2013}
M.~Gronemann, M.~J{\"{u}}nger,
  \href{http://dx.doi.org/10.1007/978-3-642-36763-2{\_}38}{Drawing clustered
  graphs as topographic maps}, in: W.~Didimo, M.~Patrignani (Eds.), Lecture
  Notes in Computer Science vol. 7704, Springer Berlin Heidelberg, Berlin,
  Heidelberg, 2013, Ch. Graph Draw, pp. 426--438.
\newblock \href {http://dx.doi.org/10.1007/978-3-642-36763-2_38}
  {\path{doi:10.1007/978-3-642-36763-2_38}}.
\newline\urlprefix\url{http://dx.doi.org/10.1007/978-3-642-36763-2{\_}38}

\bibitem{Auber2013}
D.~Auber, C.~Huet, A.~Lambert, B.~Renoust, A.~Sallaberry, A.~Saulnier,
  {GosperMap}: Using a gosper curve for laying out hierarchical data, IEEE
  Transactions on Visualization and Computer Graphics 19~(11) (2013)
  1820--1832.
\newblock \href {http://dx.doi.org/10.1109/TVCG.2013.91}
  {\path{doi:10.1109/TVCG.2013.91}}.

\bibitem{Schulz2017}
C.~Schulz, A.~Nocaj, J.~Goertler, O.~Deussen, U.~Brandes, D.~Weiskopf,
  Probabilistic graph layout for uncertain network visualization, IEEE
  Transactions on Visualization and Computer Graphics 23~(1) (2017) 531--540.
\newblock \href {http://dx.doi.org/10.1109/TVCG.2016.2598919}
  {\path{doi:10.1109/TVCG.2016.2598919}}.

\bibitem{Gansner2010}
E.~R. Gansner, Y.~Hu, S.~Kobourov, {GMap}: Visualizing graphs and clusters as
  maps, in: 2010 IEEE Pacific Visualization Symposium (PacificVis), IEEE
  Computer Society, Los Alamitos, CA, USA, 2010, pp. 201--208.
\newblock \href {http://dx.doi.org/10.1109/PACIFICVIS.2010.5429590}
  {\path{doi:10.1109/PACIFICVIS.2010.5429590}}.

\bibitem{Blades1998}
M.~Blades, J.~Blaut, Z.~Darvizeh, S.~Elguea, S.~Sowden, D.~Soni, C.~Spencer,
  D.~Stea, R.~Surajpaul, D.~Uttal,
  \href{http://dx.doi.org/10.1111/j.0020-2754.1998.00269.x}{A cross-cultural
  study of young children's mapping abilities}, Transactions of the Institute
  of British Geographers 23~(2) (1998) 269--277.
\newblock \href {http://dx.doi.org/10.1111/j.0020-2754.1998.00269.x}
  {\path{doi:10.1111/j.0020-2754.1998.00269.x}}.
\newline\urlprefix\url{http://dx.doi.org/10.1111/j.0020-2754.1998.00269.x}

\bibitem{Kairam2015}
S.~Kairam, N.~H. Riche, S.~Drucker, R.~Fernandez, J.~Heer,
  \href{http://dx.doi.org/10.1111/cgf.12642}{Refinery: Visual exploration of
  large, heterogeneous networks through associative browsing}, Computer
  Graphics Forum 34~(3) (2015) 301--310.
\newblock \href {http://dx.doi.org/10.1111/cgf.12642}
  {\path{doi:10.1111/cgf.12642}}.
\newline\urlprefix\url{http://dx.doi.org/10.1111/cgf.12642}

\bibitem{Khazaei2017}
T.~Khazaei, O.~Hoeber,
  \href{https://doi.org/10.1007/s00799-016-0170-x}{Supporting academic search
  tasks through citation visualization and exploration}, International Journal
  on Digital Libraries 18~(1) (2017) 59--72.
\newblock \href {http://dx.doi.org/10.1007/s00799-016-0170-x}
  {\path{doi:10.1007/s00799-016-0170-x}}.
\newline\urlprefix\url{https://doi.org/10.1007/s00799-016-0170-x}

\bibitem{PVPC2015}
P.~C.-I. Pang, K.~Verspoor, J.~Pearce, S.~Chang,
  \href{http://doi.acm.org/10.1145/2838739.2838772}{Better health explorer:
  Designing for health information seekers}, in: Proceedings of the Annual
  Meeting of the Australian Special Interest Group for Computer Human
  Interaction, OzCHI '15, ACM, New York, NY, USA, 2015, pp. 588--597.
\newblock \href {http://dx.doi.org/10.1145/2838739.2838772}
  {\path{doi:10.1145/2838739.2838772}}.
\newline\urlprefix\url{http://doi.acm.org/10.1145/2838739.2838772}

\bibitem{Skupin2013}
A.~Skupin, J.~R. Biberstine, K.~Börner,
  \href{https://doi.org/10.1371/journal.pone.0058779}{Visualizing the topical
  structure of the medical sciences: A self-organizing map approach}, PLOS ONE
  8~(3) (2013) 1--16.
\newblock \href {http://dx.doi.org/10.1371/journal.pone.0058779}
  {\path{doi:10.1371/journal.pone.0058779}}.
\newline\urlprefix\url{https://doi.org/10.1371/journal.pone.0058779}

\bibitem{PHMV2016}
P.~C.-I. Pang, M.~Harrop, K.~Verspoor, J.~Pearce, S.~Chang,
  \href{http://doi.acm.org/10.1145/3010915.3011843}{{What Are Health Website
  Visitors Doing: Insights from Visualisations Towards Exploratory Search}},
  in: Proceedings of the 28th Australian Conference on Computer-Human
  Interaction, OzCHI '16, ACM, New York, NY, USA, 2016, pp. 631--633.
\newblock \href {http://dx.doi.org/10.1145/3010915.3011843}
  {\path{doi:10.1145/3010915.3011843}}.
\newline\urlprefix\url{http://doi.acm.org/10.1145/3010915.3011843}

\bibitem{PCC2017}
P.~C.-I. Pang, O.~Clavisi, S.~Chang, {Engaging Consumers with Musculoskeletal
  Conditions in Health Research: A User-Centred Perspective}, Studies in Health
  Technology and Informatics 239 (2017) 104--111.
\newblock \href {http://dx.doi.org/10.3233/978-1-61499-783-2-104}
  {\path{doi:10.3233/978-1-61499-783-2-104}}.

\bibitem{saldana2015coding}
J.~Salda{\~{n}}a, {The coding manual for qualitative researchers}, Sage,
  Thousand Oaks, CA, 2015.

\bibitem{PRMB2017}
P.~C.-I. Pang, R.~P. Biuk-Aghai, M.~Yang, B.~Pang,
  \href{http://www.sciencedirect.com/science/article/pii/S1045926X17300290}{Creating
  realistic map-like visualisations: Results from user studies}, Journal of
  Visual Languages {\&} Computing 43 (2017) 60 -- 70.
\newblock \href {http://dx.doi.org/https://doi.org/10.1016/j.jvlc.2017.09.002}
  {\path{doi:https://doi.org/10.1016/j.jvlc.2017.09.002}}.
\newline\urlprefix\url{http://www.sciencedirect.com/science/article/pii/S1045926X17300290}

\bibitem{Pang:2016}
P.~C.-I. Pang, R.~P. Biuk-Aghai, M.~Yang,
  \href{http://doi.acm.org/10.1145/2968220.2968239}{What makes you think this
  is a map?: Suggestions for creating map-like visualisations}, in: Proceedings
  of the 9th International Symposium on Visual Information Communication and
  Interaction, VINCI '16, ACM, New York, NY, USA, 2016, pp. 75--82.
\newblock \href {http://dx.doi.org/10.1145/2968220.2968239}
  {\path{doi:10.1145/2968220.2968239}}.
\newline\urlprefix\url{http://doi.acm.org/10.1145/2968220.2968239}

\end{thebibliography}
